\documentclass[aps,pra,twocolumn,superscriptaddress,noeprint]{revtex4-2}
\usepackage{amsmath,amssymb,amsfonts}
\usepackage{amsthm,color}
\newcommand{\lp}[1]{\textcolor{DarkOrchid}{#1}}

\usepackage[dvipsnames]{xcolor}
\usepackage{graphicx}
\usepackage{dcolumn}
\usepackage{bm}
\usepackage{bbm}

\usepackage{lineno}

\colorlet{mylinkcolor}{teal}
\colorlet{mycitecolor}{teal}
\colorlet{myurlcolor}{teal}
\usepackage{hyperref}
\hypersetup{
  linkcolor  = mylinkcolor,
  citecolor  = mycitecolor,
  urlcolor   = myurlcolor,
  colorlinks = true,
  breaklinks = true
}
\usepackage{orcidlink}

\theoremstyle{plain}

\theoremstyle{definition}

\usepackage{cjhebrew}
\DeclareMathOperator{\Tr}{Tr}
\DeclareMathOperator{\sinc}{sinc}
\DeclareMathOperator{\Se}{Re}
\DeclareMathOperator{\Em}{Im}
\newcommand{\id}{\mathbbm{1}}

\begin{document}
\title{OH molecule as a quantum probe to jointly estimate electric and magnetic fields}
\author{Luca Previdi}
\affiliation{%
 Dipartimento di Fisica e Astronomia, Università di Bologna, I-40126 Bologna, Italy
}%
\affiliation{Dipartimento di Fisica, 
Universit\`a di Milano, I-20133 Milano, Italia}%
\author{Francesco Albarelli\,\orcidlink{0000-0001-5775-168X}}
\affiliation{%
  Scuola Normale Superiore, I-56126 Pisa, Italy
}%
\affiliation{Università di Parma, Dipartimento di Scienze Matematiche, Fisiche e Informatiche, I-43124 Parma, Italy}
\affiliation{INFN—Sezione di Milano-Bicocca, Gruppo Collegato di Parma, I-43124 Parma, Italy}

\author{Matteo G. A. Paris\,\orcidlink{0000-0001-7523-7289}}%
 \affiliation{Dipartimento di Fisica, 
Universit\`a di Milano, I-20133 Milano, Italia}%
\date{\today}

\begin{abstract}
The hydroxyl radical, hereafter referred to as the OH molecule (OHM), carries both electric and magnetic dipole moments and, as a diatomic molecule, admits a comparatively simple and accurate model.
This makes it a natural quantum probe for the joint estimation of electric and magnetic fields.
Here we study simultaneous estimation of both fields using the tools of multiparameter quantum estimation theory, explicitly accounting for the performance loss caused by measurement incompatibility.
We analyze and optimize both stationary and dynamical estimation strategies.
In the stationary regime we consider ground and thermal states of the Stark–Zeeman Hamiltonian and identify optimal operating points.
For thermal probes we find a nontrivial multiparameter effect: increasing the temperature can reduce the overall estimation error by weakening parameter correlations.
In the dynamical regime we study both pure and thermal initial states, illustrating nontrivial manifestations of incompatibility for mixed probes.
Finally, we show that an optimal sequential control protocol can overcome limitations due to noncommutativity, and we assess its robustness in the multiparameter setting.
\end{abstract}
\maketitle
\section{Introduction}\label{intro}

The OHM in its ground state has been widely studied in many fields such as ultracold chemistry~\cite{PhysRevA.71.022709,PhysRevLett.90.043006,PhysRevLett.101.203203,PhysRevA.82.022704}, precision measurements~\cite{PhysRevLett.96.143004,PhysRevA.80.022118}, and quantum computing~\cite{PhysRevA.74.061402}.
The reason for this widespread interest is that the OHM carries both electric and magnetic dipole moments, while having a diatomic structure that simplifies its modeling.
Indeed, Lara et al.~\cite{PhysRevA.78.033433} 
developed a suitable basis for writing explicitly an effective Hamiltonian for ground state of diatomic molecules under the action of static electric and magnetic fields.
The symmetries of the Hamiltonian of the OHM were explicitly studied, leading to an analytical form of the eigenvalues~\cite{PhysRevA.88.012503}.

This paper aims to discuss strategies for estimating~\cite{Helstrom1969QuantumDA,Holevo2011b,paris2009quantum} the electric and magnetic fields acting on the molecule, that is, using the OHM as a sensitive quantum probe~\cite{Giovannetti_2011,Degen_2017} to determine their values in various configurations.
Our goal is not to propose a complete experimental sensing architecture, but rather to establish quantum-limited benchmarks for a concrete Stark--Zeeman molecular system and to identify the regimes where simultaneous electric- and magnetic-field estimation is most favorable.

The estimation of a magnetic vector field with one or multiple spins is a paradigmatic problem in multiparameter quantum metrology, studied from various perspectives~\cite{Vaneph2012,Baumgratz2015,Yuan_2016,Hou2020,Gorecki2022b,Kaubruegger2023,Baamara2023,Hayashi2024,guo2023assisted}.
From an abstract point of view, the problem can be understood as estimating a unitary in SU(2) and incompatibility issues arise since the three parameters are encoded by non-commuting generators.
The more general SU(D) case has also been addressed theoretically~\cite{Kahn2007,Imai2007,Kura2017}.
Differently, our goal is to estimate electric and magnetic fields simultaneously with a single molecular probe.
Despite recent experimental effort to realize quantum sensors for both electric and magnetic fields~\cite{Esat2024}, such a scenario remains largely unexplored.
Indeed, quantum sensors are usually studied and engineered separately for the two cases~\cite{Degen_2017}.
The physical reason why this joint-estimation problem is natural for the OHM is that the same low-energy manifold carries both electric and magnetic dipole moments, so the two fields are encoded in a single probe rather than in two independent sensors.
Evidently, this application requires the probing system to be necessarily more complex, and the problem lacks the symmetry of multiparameter magnetometry.
In this respect, the OHM provides one of the simplest molecular platforms supporting simultaneous linear Stark and Zeeman responses: a near-degenerate opposite-parity doublet for electric-field sensitivity and an unpaired-electron
magnetic moment for magnetic-field sensitivity.


Indeed, there is an almost complete cancellation between the orbital and spin magnetic dipole moments in ${}^2\Pi_{1/2}$ diatomic molecules, so we instead focus on ${}^2\Pi_{3/2}$ configurations~\footnote{For clarity, we briefly recall the spectroscopic term-symbol notation for diatomic molecules.
The symbol $\Pi$ denotes $\Lambda=1$, the projection of the electronic orbital angular momentum on the internuclear axis.
The superscript $2$ gives the spin multiplicity $2S+1$ (here a doublet with $S=1/2$ for an open-shell radical with one unpaired electron).
The subscript ($1/2$ or $3/2$) denotes $\Omega$, the projection of the total electronic angular momentum on the internuclear axis~\cite{Brown2003a}.}, such as the OHM, albeit at the cost of increasing the dimensionality of the Hilbert space~\cite{PhysRevA.78.033433}.



In this paper, we use this molecular setting to assess whether, and in what sense, multiparameter incompatibility is operationally relevant for the joint estimation of electric and magnetic fields.
As we will show, incompatibility indicators can be nonzero, but their practical impact depends strongly on the probe state, parametrization, and available measurement resources.
To the best of our knowledge, the OHM probing performance has been so far examined mainly through standard spectroscopic tools based on the energy spectrum, without a systematic analysis of the full quantum Fisher information, including the contributions associated with parameter-dependent eigenstates.
We also study the efficacy of employing an adaptive control scheme to mitigate the limitations imposed by noncommutativity in specific configurations.

The paper is structured as follows.
In Sec.~\ref{multipar_QET} we give a brief review of local multiparameter quantum estimation theory, discussing in detail attainability issues related to measurement incompatibility~\cite{Albarelli_2020,Liu_2019,Demkowicz-Dobrzanski2020,Razavian_2020,PhysRevA.100.032104,Matsumoto_2002,PhysRevLett.119.130504}, and we also introduce a figure of merit that generalizes the signal-to-noise ratio (SNR) to multiple parameters in Sec.~\ref{subsec:mSNR}.
Section~\ref{OH} summarizes analytical results regarding the description of the OHM in static electric and magnetic fields, which mainly concern the choice of the basis~\cite{PhysRevA.78.033433} and the writing of the Hamiltonian~\cite{PhysRevA.88.012503}.

In Section~\ref{static}, we discuss in detail an analysis of the maximum precision attainable in the estimation of the parameters for static configurations that are closer to those commonly considered in molecular experiments.
This analysis also highlights a nontrivial finite-temperature tradeoff: thermal population of excited states can reduce parameter correlations and improve the overall multiparameter precision in some regimes, despite the accompanying loss of purity.
We will see in particular that asymptotic incompatibility measures do not always faithfully quantify the finite-copy performance loss in the parametrization of interest.

In Section~\ref{dynamical}, we focus on dynamical probes i.e. systems which evolve with time, employing evolution as a resource.
In this framework, we suggest the possibility of implementing sequential feedback schemes to bring the variance of the estimators of the parameters back to a quadratic temporal scaling.
Indeed, contrary to naive expectations, without control operations the variance is not strictly decreasing in time for this class of estimation problems.
Furthermore, we characterize the behavior and robustness of the adaptive scheme in a more realistic scenario, where the controls are implemented using an imperfect estimate of the parameters.
Finally, in Section~\ref{conclusions}, we conclude the paper with some observations and remarks.

\section{Multiparameter local quantum estimation theory}\label{multipar_QET}

This section is aimed at making this manuscript as self-contained as possible.
However, since the field of multiparameter quantum estimation is vast, rather complex and also quickly evolving, we refer the interested reader to Ref.~\cite{Albarelli_2020} for a more extended overview, to Ref.~\cite{Demkowicz-Dobrzanski2020} for an in-depth review including derivations, and to Ref.~\cite{Pezze2025} for a review of more recent advances.

\subsection{Matrix Cramér--Rao bounds}

\subsubsection{Classical bound}
Let us suppose to have a parametric family of states $ \rho_{\boldsymbol{\lambda} }$ dependent on a vector of $d$ real parameters $ \boldsymbol{\lambda } = (\lambda_1, \dots , \lambda_d)^T \in \mathbb{R}^d $.
In order to estimate these parameters, we have to perform a measurement, mathematically described by a positive operator-valued measure (POVM) $\hat{\Pi} = \{ \hat{\Pi}_k \, | \, \hat{\Pi}_k \succeq 0,\, \sum_k \hat{\Pi}_k = \id \}$.
Then we can associate a probability $p(k|\boldsymbol{\lambda})$ to the outcome $k$ of the measurement through Born rule: $p(k|\boldsymbol{\lambda})= \Tr[\rho_{\boldsymbol{\lambda} } \hat{\Pi}_k]$.
To obtain an estimate of the parameters from the measurement outcomes, we need an estimator $\tilde{\boldsymbol{\lambda}}(k)$: a function that maps measurement outcomes to potential parameter values, e.g. the maximum likelihood estimator.     
For multiple parameters, the estimation error can be quantified by the mean square error matrix of the estimator:
\begin{equation}
    V(\boldsymbol{\lambda},\{ \hat{\Pi}_k\} ) = \sum_k p(k| \boldsymbol{\lambda}  ) \big( \tilde{\boldsymbol{\lambda}}(k) -\boldsymbol{\lambda }\big) \big( \tilde{\boldsymbol{\lambda}}(k) -\boldsymbol{\lambda }\big)^T\,.
\end{equation}
We further assume the estimator to be locally unbiased, i.e. unbiased in the neighborhood of the true value of the parameter, so that $\sum_k p(k| \boldsymbol{\lambda}  ) \tilde{\boldsymbol{\lambda}}(k) = \boldsymbol{\lambda}$, and therefore $V(\boldsymbol{\lambda},\{ \hat{\Pi}_k\} )$ corresponds to the covariance matrix (CM).

We then introduce the Fisher information matrix (FIM), whose elements are defined as:
\begin{align}\label{fim}
    F_{\mu \nu}(\boldsymbol{\lambda},\{\hat\Pi_k\} ) \equiv& \sum_k \frac{\partial_{\mu} p(k|\boldsymbol{\lambda}) \partial_{\nu} p(k| \boldsymbol{\lambda})}{p(k| \boldsymbol{\lambda})}\,.
\end{align}
where $\partial_{\mu} \equiv \frac{\partial}{\partial \lambda_{\mu}}$.
The FIM quantifies the information about the parameters contained in the probability distribution.
Under the (local) unbiasedness assumption on the estimator, one can derive the Cramér--Rao bound (CRB) \cite{cramer1999mathematical,Lehmann1998theor}
for the covariance matrix:
\begin{equation}\label{CRB}
    V(\boldsymbol{\lambda},\{ \hat\Pi_k\} ) \succeq \frac{1}{M}F(\boldsymbol{\lambda},\{\hat\Pi_k\} )^{-1}\,,
\end{equation}
where $M$ is the number of identical and independent repetitions of the experiment, and we use the symbol $\succeq$ for the Loewner order, meaning that the difference between the left-hand and the right-hand sides is a positive semidefinite matrix.
Although local unbiasedness is a technical condition, in regular models the CRB in Eq.~\eqref{CRB} provides an asymptotic lower bound on the CM of regular, consistent estimators as $M\to\infty$. Moreover, under standard conditions the maximum likelihood estimator is asymptotically efficient and attains this bound.

\subsubsection{Quantum bound}

Clearly, the Born rule implies a dependence of the information on the chosen POVM, consequently, one would like to maximize the FIM in Eq.~\eqref{fim} or minimize its inverse, i.e. the right-hand side of Eq.~\eqref{CRB}, over the set of possible measurements.
While these optimizations are equivalent and lead to a simple result for a single parameter, in the multiparameter case the situation is more involved.
Indeed, in general, it is not even possible to optimize a matrix in the Loewner sense. However, it is possible to derive matrix upper bounds on the FIM of any POVM that only depend on the parametric family of states.

The most well-known quantum generalization of the FIM is based on the symmetric logarithmic derivative (SLD) operators $\hat{L}_{\mu}$, defined implicity by the equation
\begin{eqnarray}\label{sldapp}
     \partial_{\mu } \rho_{\boldsymbol{\lambda}} &=& \frac{ \hat{L}_{\mu} \rho_{\boldsymbol{\lambda}}+ \rho_{\boldsymbol{\lambda}} \hat{L}_{\mu} }{2}\,,
\end{eqnarray}
and it is usually known simply as the quantum Fisher information matrix (QFIM)~\cite{Helstrom1967,paris2009quantum,Liu_2019}, defined as:
\begin{eqnarray}
    Q_{\mu \nu}(\boldsymbol{\lambda})& \equiv & \Tr \bigg[\rho_{\boldsymbol{\lambda}}  \frac{\hat{L}_{\mu} \hat{L}_{\nu}+ \hat{L}_{\nu} \hat{L}_{\mu}}{2} \bigg] \label{QFIM} \,.
\end{eqnarray}
As mentioned, one can prove the matrix inequality $Q_{\mu \nu}(\boldsymbol{\lambda}) \succeq F(\boldsymbol{\lambda},\{\hat\Pi_k\} )$ for any POVM, which in turn implies the so-called quantum Cramér-Rao bound (QCRB) on the CM:
\begin{equation}\label{bound}
    V(\boldsymbol{\lambda},\{\hat\Pi_k\}) \succeq \frac{1}{M}F(\boldsymbol{\lambda},\{\hat\Pi_k\})^{-1} \succeq\frac{1}{M}Q(\boldsymbol{\lambda})^{-1}\,.
\end{equation}
Writing the probe state in its diagonal form $ \rho_{\boldsymbol{\lambda}} = \sum_n r_n |\psi_n \rangle \langle \psi_n|$, the explicit expression of the QFIM Eq.\;(\ref{QFIM}) is given by
\begin{align}\label{QFIMexpl}
    &Q_{\mu \nu}(\boldsymbol{\lambda}) =  \notag \sum_n \frac{\partial_{\mu} r_n \partial_{\nu} r_n}{r_n}+
    \sum_{n \neq m} \frac{(r_n-r_m)^2}{r_n+r_m}\times \\&  \big[\langle \psi_n | \partial_{\mu} \psi_m \rangle \langle \partial_{\nu} \psi_m| \psi_n \rangle +\langle \psi_n | \partial_{\nu} \psi_m \rangle \langle \partial_{\mu} \psi_m| \psi_n \rangle \big]\,,
\end{align}
notice that both eigenvalues $r_n$ and eigenstates $| \psi_n \rangle$ depend on the parameters $\boldsymbol{\lambda}$, but we drop the explicit dependence for a more compact notation.
For pure-state models $\rho_{\boldsymbol{\lambda}} = | \psi \rangle \langle \psi|$, Eq.~\eqref{QFIMexpl} simplifies to
\begin{equation}\label{QFIpure}
     Q_{\mu \nu}(\boldsymbol{\lambda}) = 4 \Se \big[\langle \partial_{\mu}\psi |\partial_{\nu} \psi \rangle + \langle \partial_{\mu}\psi | \psi \rangle\langle \partial_{\nu}\psi | \psi \rangle \big]\,.
\end{equation}

\subsection{Scalar quantum Cramér-Rao bounds}

Since a matrix-valued optimization is not always well-posed, it is common to focus on a scalar figure of merit, which can be unambiguously compared for different POVMs and thus optimized.
In particular, the standard choice is to consider the weighted trace of the CM $\Tr \left[ W \, V(\boldsymbol{\lambda},\{\hat\Pi_k\} ) \right]$, by introducing a weight matrix $W >0$.
In practice, this quantity represents the sum of variances of linear combinations of the parameters.
For the simplest choice $W=\id_d$ this is just the sum of variances in the original parametrization.

\subsubsection{SLD and Holevo--Cramér--Rao bounds}

Starting from the matrix bound Eq.~\eqref{bound}, we obtain:
\begin{eqnarray}
     \Tr \left[ W \, V(\boldsymbol{\lambda},\{\hat\Pi_k\} ) \right] &\ge& \frac{1}{M} C^S(\boldsymbol{\lambda}, W)\,,
\end{eqnarray}
where we have introduced
\begin{eqnarray}
    C^S(\boldsymbol{\lambda}, W) &\equiv& \Tr[W Q^{-1}] \,,\label{scalar}
\end{eqnarray}
which we call the SLD scalar bound.
Since in this work, we will mostly use the identity as the weight matrix, we simplify the notation as $C^S(\boldsymbol{\lambda}) \equiv C^S(\boldsymbol{\lambda}, \mathbb{I})$.

A tighter lower bound on the scalar estimation error, known as Holevo--Cramér--Rao bound (HCRB), can be formulated as~\cite{Holevo2011b,HOLEVO1977251,Nagaoka1989}:
\begin{equation} \label{holevo}
    M\Tr[WV] \ge C^H(\boldsymbol{\lambda}, W) \ge  C^S(\boldsymbol{\lambda}, W)\,,
\end{equation}
with $C^H(\boldsymbol{\lambda}, W)$ defined through the following minimization:
\begin{align}\label{CH}
      \notag C^H(\boldsymbol{\lambda}, W) \equiv& \min_{U \in \mathbb{S}^d, X \in \mathbb{X}_{\lambda}} \left\{  \Tr[W U ] \, | \, U \succeq Z[\hat{\boldsymbol{X}}] \right\} \\ =& \notag  \min_{ \hat{\boldsymbol{X}} \in \mathbb{X}_{\lambda}} \big\{  \Tr\left[W \Se\left[ Z[\hat{\boldsymbol{X}}] \right] \right] \\ &+ \left \lVert \sqrt{W} \Em \left[ Z[\hat{\boldsymbol{X}}] \right] \sqrt{W} \right\rVert_1   \big\} \,, 
\end{align} 
where $ \lVert A \rVert_1 = \Tr\left[ \sqrt{A^{\dag} A} \right]$ is the trace norm of $A$, $\mathbb{S}^d$ denotes the space of $d$-dimensional real symmetric matrices and $Z[\hat{\boldsymbol{X}}]$ is the Hermitian (complex) $d \times d$ matrix with elements
\begin{equation}
    Z_{\mu \nu}[\hat{\boldsymbol{X}}] = \Tr[\rho_{\boldsymbol{\lambda }} \hat X_{\mu} \hat X_{\nu}]\,,
\end{equation}
with the collection of operators $X$ belonging to the set:
\begin{equation}
       \mathbb{X}_{\boldsymbol{\lambda} } = \{ \hat{\boldsymbol{X}} = ( \hat X_1, \dots, \hat X_d) \, | \, \Tr[\partial_{\mu} \rho_{\boldsymbol{\lambda}} \hat X_{\nu}] = \delta_{\mu \nu} \}\,.
\end{equation}

\subsubsection{Asymptotic and single-copy attainability}

The classical CRB is attainable in the limit $M \to \infty$, i.e. for a large number of repetitions.
Since an independent copy of the state must be prepared for each experiment repetition, the whole procedure uses $M$ copies of the state, i.e. the overall state is $\rho_{\boldsymbol{\lambda}}^{\otimes M}$.
Crucially, in the quantum case it is theoretically possible to measure all the copies collectively.
In this scenario, it has been shown that the HCRB in Eq.~\eqref{holevo} is optimal and saturable~\cite{yamagata2013quantum, hayashi2008asymptotic,yang2019attaining}.
Thus, the HCRB represents the fundamental asymptotic limit imposed by quantum mechanics on the estimation precision when many independent copies of the state are available to be measured.

In practice, since we have introduced a scalar figure of merit, one can also consider the so-called most informative (or tight) bound, obtained by minimizing it over POVMs $\min_{ \{ \hat\Pi_k \} } \Tr \left[ W \, F(\boldsymbol{\lambda},\{\hat\Pi_k\} )^{-1} \right]$.
This quantity represents the best scalar error achievable when only measurement on a single-copy of the state can be performed, but the experiment is repeated many times~\cite{Hayashi2023a}.
Unfortunately, no effective methods to compute this quantity are known for Hilbert spaces of dimension larger than 2, and this has been related to the hardness of characterizing the cone of separable operators~\cite{Hayashi2023a}.
Clearly, the most informative bound is generally larger than the HCRB, since single-copy measurements are strictly less powerful than collective ones.
Tighter bounds exist for this scenario, such as the Nagaoka--Hayashi bound~\cite{Conlon2020}, but it is unclear when they are attainable and we will not consider them in this work, focusing instead on the fundamental HCRB.

The case of pure states is special: the HCRB is always attainable with single-copy measurements, and thus coincides with the most informative bound~\cite{Matsumoto_2002}.



\subsection{Measurement incompatibility}

The fact that the SLD bound $C^S$ is not attainable is ultimately related to measurement incompatibility between the optimal measurements to estimate different parameters.
Thus, the discrepancy between $C^S$ and $C^H$ can be understood as a quantifier of incompatibility in multiparameter estimation~\cite{Belliardo2021}.
In this section, we review upper bounds on this gap, and the conditions under which it is zero.

\subsubsection{Upper bounds and weak commutativity}

Even if the HCRB can be efficiently evaluated through a semidefinite program (SDP)~\cite{albarelli2019evaluating}, it remains generally difficult to compute and it is useful to identify the discrepancy with $C^S$ and conditions for equality.
The HCRB is bounded from above as~\cite{carollo2019quantumness,Tsang2019}
\begin{align}
        C^S(\boldsymbol{\lambda}, W) \le  C^H(\boldsymbol{\lambda}, W)
         & \notag\le \overline{C^H} (\boldsymbol{\lambda}, W) \\ & \le C^S(\boldsymbol{\lambda}, W)(1+R)\, ,
         \label{R_bound}
\end{align}
where the weight-dependent bound~\cite{Tsang2019,Albarelli_2020} $\overline{C^H}$ and the weight-independent \textit{asymptotic incompatibility}~\cite{carollo2019quantumness,Razavian_2020,candeloro2021properties,candeloro2024dimension} $R$  are defined in terms of the mean Uhlmann curvature (UC) matrix $D$~\cite{UHLMANN1986229,Dittmann_1999,carollo2017uhlmann,Albarelli_2022}
\begin{equation} \label{uc}
    D_{\mu \nu } = -\frac{i}{2} \Tr\big[\rho_{\boldsymbol{\lambda} }[\hat{L}_{\mu}, \hat{L}_{\nu}] \big]\,,
\end{equation}
as
\begin{equation}\label{ch_bar}
    \overline{C^H} (\boldsymbol{\lambda}, W) = C^S (\boldsymbol{\lambda}, W)+\lVert \sqrt{W} Q^{-1} D Q^{-1} \sqrt{W} \rVert_1 \, ,
\end{equation}
and
\begin{equation} \label{R}
    R = \lVert i Q^{-1} D \rVert_{\infty}\,.
\end{equation}
with $\lVert A \rVert_{\infty}$ being the maximum eigenvalue of $A$.

The condition:
\begin{equation}
    D = 0\,,
\end{equation}
which clearly implies $R = 0$, is sometimes referred to as the weak commutativity condition (WCC).
While from Eq.~\eqref{R_bound} it is immediate that the WCC $D=0$ (thus $R=0$) is sufficient for the equality $C^S(\boldsymbol{\lambda}, W) = C^H(\boldsymbol{\lambda}, W)$ for all $W$, it can be shown that this is also a necessary condition~\cite{Ragy_2016}.
Thanks to this, $R$ can be considered a bonafide quantifier of asymptotic incompatibility.
This quantity is invariant under reparametrizations of the model, and can thus be understood as a geometrical feature of the model.
However, this also means that this quantity captures incompatibility in the worst-case scenario.
For a fixed parametrization of interest, the weight-dependent bound $\overline{C^H}$ is usually more informative.

\subsubsection{Attaining the QFI matrix}

Instead of focusing on scalar quantities, we now focus on the conditions for the existence of a POVM $\{ \hat\Pi_k\}$ such that $F(\boldsymbol{\lambda}, \{ \hat\Pi_k\}) = Q(\boldsymbol{\lambda})$, with single-copy measurements.
A necessary condition for the existence of such a POVM is known as partial commutativity condition (PCC)~\cite{PhysRevA.100.032104}, and stated as follows:
\begin{equation}
    \langle \phi| [\hat{L}_{\mu} ,\hat{L}_{\nu}] | \psi \rangle = 0 \quad \forall | \psi \rangle, | \phi \rangle \in \operatorname{supp}(\rho_{\boldsymbol{\lambda}})\,, \label{eq371}
\end{equation}
where $\operatorname{supp}(\rho)$ is the support of the state $\rho$, i.e. for finite-dimensional systems it is the subspace spanned by eigenstates corresponding to nonzero eigenvalues.
To be precise, we also need a regularity assumption that the rank of the state, i.e. the number of nonzero eigenvalues, does not change in an infinitesimal neighborhood of the true parameter value, where all quantities are evaluated.
Clearly, for a full-rank model (with no zero eigenvalues), the PCC is equivalent to asking the commutativity of the SLDs over the whole Hilbert space.
Therefore, in this scenario, the PCC is also sufficient since the optimal POVM can be chosen to be the projectors on the common eigenstates of the SLDs.

\subsubsection{Compatibility conditions for pure-state models}

For pure-state models, the WCC reduces to the PCC in Eq.~\eqref{eq371} and it has been proven to be sufficient~\cite{Matsumoto_2002} for finding a POVM that attains an equality between the QFIM and FIM. 
Moreover, it can also be reworked in a more constructive manner~\cite{PhysRevLett.119.130504}.

Indeed, given a projection-valued measure (PVM) consisting of projectors that are not orthogonal to the probe $|\pi_k \rangle \langle \pi_k |$, and projectors orthogonal to the probe $|\theta_k \rangle \langle \theta_k |$, it saturates the QCRB if and only if:
\begin{equation}\label{pezzeort1}
        \lim_{\alpha \rightarrow \lambda} \frac{\Em\big[\langle \partial_{\mu} \psi_{\alpha}| \theta_k \rangle \langle \theta_k | \psi_{\alpha} \rangle \big]}{| \langle \psi_{\alpha} | \theta_k \rangle|} = 0
\end{equation}
and
\begin{equation}\label{pezzeort}
        \Em\big[\langle \partial_{\mu} \psi_{\lambda}| \pi_k \rangle \langle \pi_k |   \psi_{\lambda} \rangle \big] = |\langle \psi_{\lambda} | \pi_k \rangle|^2 \Em\big[\langle \partial_{\mu} \psi_{\lambda} | \psi_{\lambda } \rangle\big]\,,
\end{equation}
$\forall \mu ,k$.
Assuming that there exists $\nu$ such that $\langle \partial_{\nu} \psi_{\lambda}| \theta_k \rangle \neq 0$, the condition (\ref{pezzeort1}) becomes:
\begin{equation}
    \Em[\langle \partial_{\mu} \psi_{\lambda}| \theta_k \rangle \langle \theta_k |  \partial_{\nu} \psi_{\lambda} \rangle ] = 0\,.
\end{equation}

Consider now the scenario in which the probe $|\psi_\lambda \rangle$ and its derivatives with respect to the parameters $| \partial_\mu \psi_\lambda \rangle$ can be expressed as a real linear combination on a given basis.
Meaning that, given a basis of the Hilbert space $\{|b_k \rangle\}_k$, we have:
\begin{equation} \label{RPDcond}
    |\psi \rangle = \sum_k c_k | b_k \rangle \;\text{and}\; | \partial_\mu \psi \rangle = \sum_k d^{\mu}_k | b_k \rangle \; 
\end{equation}
with $c_k, d^{\mu}_k \in \mathbb{R}$ $\forall k, \mu$.
Then it is straightforward to notice that any PVM $| \pi_k \rangle \langle \pi_k | $ such that:
\begin{equation} \label{RPMcond}
    |\pi_k \rangle = \sum_j f_j^k | b_k\rangle \; \text{with} \; f_j^k \in \mathbb{R} \; \forall j,k\,,
\end{equation}
satisfies the condition Eq.~\eqref{pezzeort}. Indeed, we will have:
\begin{equation}
    \Em\big[\langle \partial_{\mu} \psi_{\lambda} | \psi_{\lambda } \rangle\big] = 0\,,
\end{equation}
and the condition in Eq.~\eqref{pezzeort} becomes:
\begin{equation}\label{less}
    \Em\big[\langle \partial_{\mu} \psi_{\lambda}| \pi_k \rangle \langle \pi_k |   \psi_{\lambda} \rangle \big] =0\,.
\end{equation}
which is automaticaly satisfied for a PVM $\{|\pi_k \rangle \langle \pi_k |\}_k$ satisfying condition Eq.~\eqref{RPMcond}.
We point out that, in this scenario, the condition Eq.~\eqref{RPMcond} it is not a necessary condition but just a sufficient one, while Eq.~\eqref{less} is a necessary and sufficient condition.
Consequently, in principle, there can exist PVMs satisfying the condition in Eq.~\eqref{less} while not satisfying the condition Eq.~\eqref{RPMcond}.
However, if the aim is to find at least a measurement, it is much easier to seek for a measurement satisfying Eq.~\eqref{RPMcond}.

\subsection{Signal-to-noise ratio for multiple parameters}

\label{subsec:mSNR}

For a single parameter, the signal-to-noise ratio (SNR) is simply the ratio between the true value of the parameter and the standard deviation of its estimator, which can be upper bounded through the CRB, and for quantum models through the QCRB~\cite{paris2009quantum}.
Unlike the Fisher information or the variance, the SNR is dimensionless and thus often useful to streamline the analysis and make some qualitative features of the problem more explicit.

Moving to multiple parameters, there is no unique way to build a single scalar quantity that summarizes the relative precision for all parameters.
In principle, one can define a matrix-valued SNR, as recently done in Ref.~\cite{mihailescu2024multiparameter}.
Here we introduce a scalar figure of merit 
that we call the multiparameter signal-to-noise ratio (mSNR), defined as:
\begin{align} 
    \text{mSNR} = & \frac{1}{\sqrt{\sum_{\mu}  \frac{V_{\mu \mu } ( \boldsymbol{\lambda},\{ \hat\Pi_k\} )}{  \lambda_{\mu}^2}}} = \left(\Tr[W_0(\boldsymbol{\lambda}) \, V(\boldsymbol{\lambda},\{ \hat\Pi_k\} ) ]\right)^{-\frac{1}{2}} \nonumber \\ 
     & \le \left[ C^S\left(\boldsymbol{\lambda}, W_0(\boldsymbol{\lambda})\right)\right]^{-\frac{1}{2}} = \overline{{\text{mSNR}}} \,, \label{snr}
\end{align}
where $W_0(\boldsymbol{\lambda}) = \operatorname{diag}\left(\frac{1}{\lambda_1^2}, \dots, \frac{1}{\lambda_d^2}\right)$ and $\overline{{\text{mSNR}}}$ is the upper bound on the mSNR induced by the QCRB.

As expected, for a single parameter the mSNR collapses to the standard signal-to-noise ratio (SNR).
However, a crucial feature of our specific definition is that the mSNR drops to zero when the variance of even just one parameter diverges, i.e. it is dominated by the largest error.
One could also introduce a different multiparameter generalization, the sum of the SNRs relative to different parameters.
However, this quantity is dominated by the smallest error not by the largest: if the variance of one estimator diverges, one term in the sum goes to zero, but not the total.
Since we care about simultaneously estimating \emph{all} the parameters $\boldsymbol{\lambda}$ with great precision, we believe the mSNR that we have introduced is the more suitable between the two.
We also stress that this definition retains the effect of correlations through the full covariance matrix entering the scalar bound.
In the applications below, the behaviour of the mSNR is similar to that of $C^S$, but this agreement is not imposed by construction: it provides a useful consistency check that the optimal working regimes are not an artefact of the particular reduction to the scalar $C^S$.











\section{OH effective Hamiltonian }
\label{OH}
    
The Stark-Zeeman Hamiltonian in the OHM ground state ${}^2 \Pi_{3/2}$~\cite{bernath2020spectra,Brown2003a} can be written as~\cite{PhysRevA.78.033433,PhysRevA.85.033427}:
\begin{equation}
    \hat{H} = \hat{H}_0 - \hat{\boldsymbol{\mu}}_e \cdot \boldsymbol{E} - \hat{\boldsymbol{\mu}}_b \cdot \boldsymbol{B}\,,
\end{equation}
where $\hat{H}_0$ represents the field-free $\Lambda$-doubling
Hamiltonian with energy gap $ \hbar \Delta_0$ \cite{PhysRevA.88.012503}, where $\Delta_0 = 2 \pi \times 1.667  $ GHz, is the lambda-doubling parameter.
Moreover, $\boldsymbol{\hat{\mu}}_e$ and $\boldsymbol{\hat{\mu}}_b$ are the electric and magnetic dipole moment operators of the OHM, and $\boldsymbol{E}$ and $\boldsymbol{B}$ are the external electric and magnetic fields applied to the OHM.
This modelling is valid when the hyperfine structure can be neglected, i.e. in the strong-field limit ($|\boldsymbol{E}|>1\,\mathrm{kV/cm}$ and $|\boldsymbol{B}|>100\,\mathrm{G}$) and/or at temperatures higher than $5\,\mathrm{mK}$~\cite{PhysRevA.78.033433}.
The zero-temperature results below should therefore be understood as the ground-state limit of this effective Hamiltonian, while the finite-temperature results are physically reliable within the quoted validity regime.

\begin{figure}
    \centering
    \includegraphics[width = 5cm]{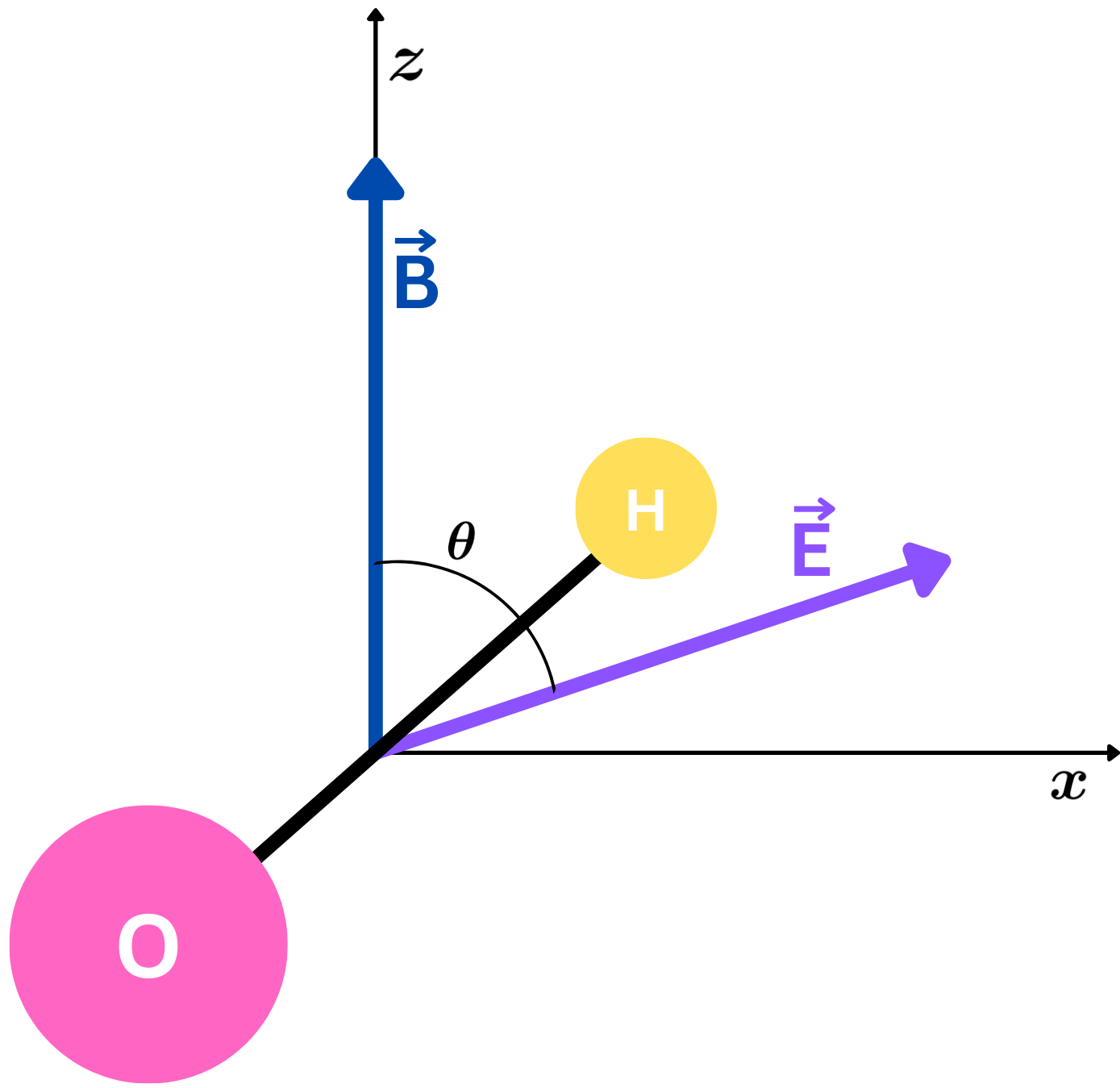}
    \caption{Pictorial representation of the OHM with static electric and magnetic fields acting on it.
    The magnetic field defines the $z$ axis, while the electric field lies in the $x$--$z$ plane and forms an angle $\theta$ with the positive $z$ axis towards the positive $x$ axis.}
    \label{scheme_oh} 
\end{figure}

Using the basis $| J, M, \Bar{\Omega}, \varepsilon \rangle$ proposed by Lara et al.~\cite{PhysRevA.78.033433}, the Hamiltonian in the OHM ground manifold ${}^2 \Pi_{3/2}$ can be described as an ${8{\times}8}$ Hermitian matrix.
This dimensionality follows from the fact that, for $J = 3/2$, the projection $M$ of the total angular momentum on the laboratory $z$ axis takes $2J+1 = 4$ values ($M = \pm 3/2, \pm 1/2$), and each such rotational level is further split into two opposite-parity components by $\Lambda$-doubling, labeled by $\varepsilon = \{e, f\}$.
To simplify the notation, we denote $\{| e_j \rangle \} = \{ | J = 3/2 , M, \Bar{\Omega} = 3/2 , \varepsilon \rangle \}$ as the canonical basis.
\lp{Where the correspondence with $j$ is given by setting $\varepsilon = e$ and increasing $M$ for $j = 1, \dots, 4$ and $\varepsilon = f$ and increasing $M$ for $j = 5, \dots, 8$. }
Here, $J = 3/2$ is the total angular momentum of the OHM, $M$ is the projection of $J$ in the laboratory reference frame,
$\Bar{\Omega}$ is the projection of $J$ on the internuclear axis (fixed to $\Bar{\Omega}=3/2$ in the ${}^2\Pi_{3/2}$ manifold),
and $\varepsilon$ denotes the $e$-$f$ symmetry.
We assume, as in Ref.~\cite{PhysRevA.78.033433}, that both the electric and magnetic dipole moments lie along the internuclear axis.
We choose the $z$ axis to be aligned with $\boldsymbol{B}$ and take $\boldsymbol{E}$ to lie in the $x$--$z$ plane, forming an angle $\theta$ with the positive $z$ axis towards the positive $x$ axis, as shown in Fig.~\ref{scheme_oh}.



Under these assumptions, the Hamiltonian reads:
\begin{align} \label{ham}
 \hat{H}(\boldsymbol{\lambda}  ) = & \notag -\Delta\, \hat{\mathcal{T}}_{300} - \frac{4}{5}\lambda_1(2\, \hat{\mathcal{T}}_{030} + \hat{\mathcal{T}}_{003})\\& \notag +\frac{2}{5} \lambda_2( 2\, \hat{\mathcal{T}}_{130} + \hat{\mathcal{T}}_{103})\\ & -\frac{2}{5}  \lambda_3(\sqrt{3}\, \hat{\mathcal{T}}_{101} + \hat{\mathcal{T}}_{111} + \hat{\mathcal{T}}_{122})\,.
\end{align}
where $\Delta = \hbar \Delta_0$, $\mu_B$ is the Bohr magneton, $\mu_e = 1.66$ D is the OHM electric dipole moment.
Moreover, $\lambda_1 = \mu_B\,|\boldsymbol{B}|$, $\lambda_2 = \mu_e\,|\boldsymbol{E}| \,\cos{\theta}$, $\lambda_3 = \mu_e\,|\boldsymbol{E}| \,\sin{\theta}$
and
\begin{equation}
    \hat{\mathcal{T}}_{ijk} = \frac{1}{2}\hat \sigma_i \otimes \hat \sigma_j \otimes \hat \sigma_k
\end{equation}
where $\sigma_i$ are the Pauli matrices.
\begin{align}
  & \notag \hat \sigma_0 = \left(
\begin{array}{cc}
 1 & 0 \\
 0 & 1 \\
\end{array}
\right), \,\,
\hat \sigma_1 = \left(
\begin{array}{cc}
 0 & 1 \\
 1 & 0 \\
\end{array}
\right),
\\ &
\hat \sigma_2 = \left(
\begin{array}{cc}
 0 & -i \\
 i & 0 \\
\end{array}
\right), \,\,
\hat \sigma_3 = 
\left(
\begin{array}{cc}
 1 & 0 \\
 0 & -1 \\
\end{array}
\right).
\end{align}

With this choice of reference frame, the estimation problem is reduced to three parameters, from which we can infer the magnitudes of the electric and magnetic fields and the relative angle between their directions.
In a fully general laboratory-frame reconstruction, the two vector fields would require more independent parameters.
Such a model would also require specifying a concrete experimental apparatus, including the fields defining the laboratory frame, trap geometry, preparation, and readout; this lies beyond the scope of the present work.
Thus, the model considered here should be regarded as a controlled theoretical benchmark for joint electric- and magnetic-field estimation, rather than as a complete experimental protocol.


The parameters of interest $\lambda_i$ have units of energy.
We will express energies in temperature units (kelvin) by dividing by the Boltzmann constant $k_B$.
This is standard in the cold-molecule literature and is equivalent to working in units where $k_B=1$.

For reference, the natural electric and magnetic field scales associated with the energy gap $\Delta$ are
\begin{equation}
E_\Delta=\frac{\Delta}{\mu_e}\simeq 199\,\mathrm{kV/m},\qquad
B_\Delta=\frac{\Delta}{\mu_B}\simeq 0.119\,\mathrm{T}.
\end{equation}
The field ranges used in the numerical plots are above the lower strong-field scales quoted above; for instance, the smallest electric field shown in Fig.~\ref{fig_gibbs}(b) is of order $10^4\,\mathrm{kV/m}$.

\section{Parameter estimation with stationary probes}\label{static}

In this section we consider \emph{static} probes for the estimation procedure, i.e. states that do not evolve in time.
Namely, states $\rho_t$ such that $\dot{\rho}_t = 0$; for an isolated system this is ensured by $[\hat{H}(\boldsymbol{\lambda}),\rho_0]=0$.
In particular, we focus on the ground and Gibbs thermal state of the parameter-dependent Hamiltonian.
Operationally, these states can be prepared by a parameter-independent protocol (cooling or waiting for equilibration), while their dependence on $\boldsymbol{\lambda}$ arises solely through $\hat{H}(\boldsymbol{\lambda})$.
We note that quantum estimation of Hamiltonian parameters encoded in thermal (or effective-thermal) states has increasingly attracted attention in recent years~\cite{Troiani2017,Garcia-Pintos2024,Abiuso2025}.
Indeed, Gibbs states often provide a useful phenomenological model for equilibrium states even in the absence of detailed microscopic knowledge of the equilibration mechanism.


We consider an estimation procedure that consists of preparing a probe and performing a measurement.
Since quantum states are rarely distinguishable with certainty~\cite{qiu2009minimumerror}, even if it is possible to associate each parameter value with a single state, it will suffer from an uncertainty due to the impossibility of distinguishing states with a single measurement.
In order to reduce this uncertainty, the procedure described above is repeated multiple times, as the CRB in Eq.~\eqref{CRB} is saturated in the asymptotic limit.
This can be done through several strategies all of which require preparing multiple copies of the probe.

\subsection{Ground state probing of aligned fields}

First, we focus on aligned fields, considering the Hamiltonian Eq.~\eqref{ham} with $\lambda_3 = 0$, namely $\hat{H}_a = \hat{H}(\lambda_1, \lambda_2 = \mu_e |\boldsymbol{E}|, \lambda_3 = 0)$.
In this case, it is possible to find an analytical form of the ground state $ |E_{a0} \rangle$ of $\hat{H}_a$:
\begin{align} \label{probe_static_aligned}
    |E_{a0} \rangle  = & \notag \frac{-5 \Delta -\sqrt{P}}{
   \sqrt{\left(5 \Delta +\sqrt{P}\right)^2+36 \lambda_2^2}}|e_1\rangle \\ &+\frac{6 \lambda_2}{
   \sqrt{\left(5 \Delta +\sqrt{P}\right)^2+36 \lambda_2^2}} |e_5 \rangle
\end{align}
for all the values of the parameters; we have introduced the parameter $P = 25 \Delta ^2+36 \lambda_2^2$.
Exploiting Eq.~\eqref{QFIpure} it is possible to explicitly evaluate the QFIM  relative to $\lambda_1$ and $\lambda_2$.
However, it is already clear from Eq.~\eqref{probe_static_aligned} that the probe does not depend on $\lambda_1$, consequently its derivatives will vanish and with them all the entries of the QFIM relative to $\lambda_1$.
This implies that the system does not carry information about the magnetic field. Nevertheless, it is possible to estimate the electric field alone.
The QFI relative to $\lambda_2$ reads:
\begin{equation}
   Q_{\lambda_2}=  \frac{900 \Delta^2 }{P}
\end{equation}
The QCRB is then attained by any measurement satisfying conditions Eqs.~\eqref{pezzeort1} and~\eqref{pezzeort}. 
Notice that in the single-parameter scenario the condition of attainability for a projector orthogonal to the probe Eq.~\eqref{pezzeort} is always satisfied~\cite{PhysRevA.100.032104}.
Moreover, since the probe in Eq.~\eqref{probe_static_aligned} and its derivatives satisfy the condition Eq.~\eqref{RPDcond}, the QCRB is saturated by any PVM satisfying condition Eq.~\eqref{RPMcond}.

\subsection{Ground state probing of generic fields}
\label{sec52}

We now move to a generic orientation for the fields, focusing on the ground state $|E_0 \rangle$ of the full Hamiltonian $\hat{H} = \hat{H}(\lambda_1, \lambda_2 , \lambda_3)$ in Eq.~\eqref{ham} as the probe.
Given the spectral decomposition of the Hamiltonian $\hat{H} = \sum_n E_n |E_n \rangle \langle E_n |$, we can exploit a generalization of the Hellman--Feynman theorem~\cite{PhysRev.56.340} to express the products $\langle \partial_{\mu} E_n | E_k \rangle$ as expectation values of Hamiltonian derivatives, avoiding the need to evaluate derivatives of the eigenstates.
More details, are given in Appendix~\ref{app:unitary_metrology}, see in particular Eq.~\eqref{hfth}.
With this approach we can write the following expression for the QFIM:
\begin{equation}\label{QFIgs}
    Q_{\mu \nu} = 4 \sum_{n= 1}^7 \frac{\Se\big[\langle E_0 |\partial_{\mu} \hat{H} | E_n \rangle\langle E_n |\partial_{\nu} \hat{H} | E_0 \rangle\big]}{\delta E_n^2} \,,
\end{equation}
where $\delta E_n = E_n-E_0$.
We thus obtain an expression of the QFIM which is independent of the derivatives of the states, simplifying considerably the numerical implementation.
\begin{figure}
    \centering
    \includegraphics[width = 8cm]{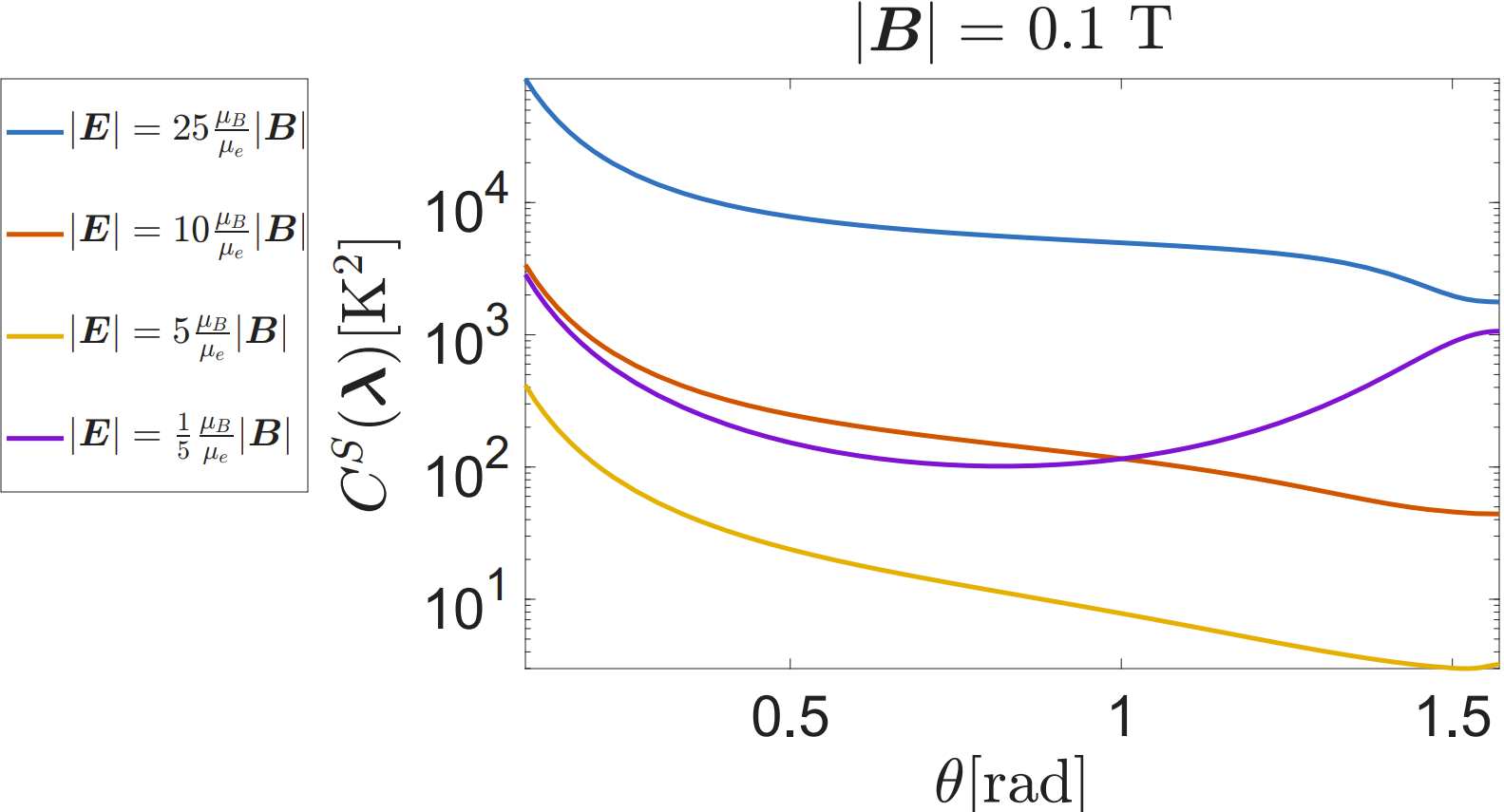}
    \caption{Scalar bound on the total estimation error $C^S( \boldsymbol{\lambda}$), in units of ${\text{kelvin}}^2$ $[\text{K}^2]$, for the ground state, plotted as a function of the angle $\theta$ between $\boldsymbol{B}$ and $\boldsymbol{E}$, and for fixed values of $| \boldsymbol{B}|$ and $|\boldsymbol{E}|$, shown in the legend.}
    \label{fig64_f} 
\end{figure}

\begin{figure*}
    \centering
    \includegraphics[width = 17cm]{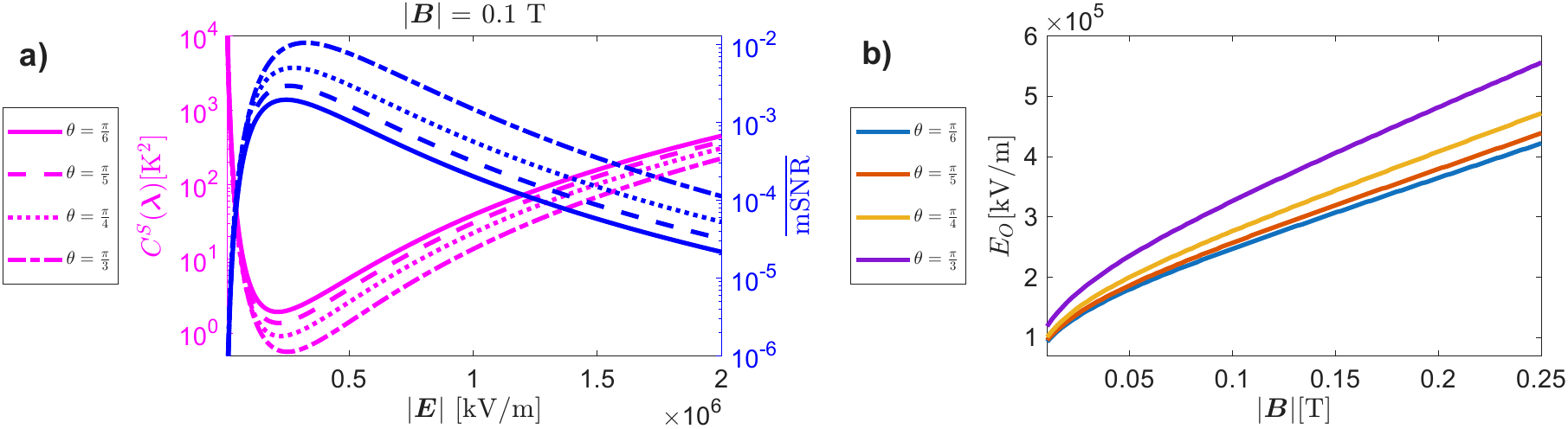}
    \caption{
    (a) Scalar bound on the total estimation error $C^S( \boldsymbol{\lambda}$), in units of kelvin${}^2\,[\text{K}^2]$,  and multiparamater quantum signal-to-noise ratio $\overline{\text{mSNR}}$ for the set of parameters $(\lambda_1,\lambda_2, \lambda_2)$ as a function of the magnitude of the electric field $|\boldsymbol{E}|$ and for fixed values of $\theta$ and $|\boldsymbol{B}|$.
     (b) Value of the electric field at the optimal working point $E_{\text{O}}$ plotted as a function of the value of the magnetic field $| \boldsymbol{B}|$ for fixed values of $\theta$.
     The optimal working point is defined as the maximum of the $\overline{\mathrm{mSNR}}$.
     }
     \label{fig_snr}
\end{figure*}

In Fig.~\ref{fig64_f} we show $C^S$, i.e. the trace of the inverse of the QFIM, as a function of $\theta$.
The plot is shown from $0$ to $\pi/2$ because of the symmetry discussed below; the corresponding behaviour in the interval $\pi/2<\theta<\pi$ is obtained by reflection around $\pi/2$.
Given the unitary $\hat{U}=2\,\hat{\mathcal{T}}_{3,0,3}$, we have $\hat{H}^\prime = \hat{U} \hat{H}(\lambda_1, \lambda_2, \lambda_3) \hat{U}^\dag = \hat{H}(\lambda_1, -\lambda_2, \lambda_3)$.
Then, $\hat{H}^\prime = \sum_n E_n |E_n^\prime \rangle \langle E_n^\prime |$ with $|E_n^\prime \rangle = \hat{U} | E_n \rangle$, and as a consequence we have:
\begin{equation}
    \langle E_i^\prime |\partial_\mu \hat{H}^\prime |E_j^\prime \rangle = \langle E_i |\partial_\mu \hat{H} |E_j \rangle
\end{equation}
and the QFI in Eq.~\eqref{QFIgs} is unchanged.
Furthermore, it is straightforward to notice that the transformation $( \lambda_1, \lambda_2, \lambda_3) \rightarrow ( \lambda_1, -\lambda_2, \lambda_3)$ is equivalent to $ \theta \rightarrow \pi- \theta$, consequently $Q_{\mu \nu}( \theta) = Q_{\mu \nu}( \pi-\theta)$.
This explains the symmetry of Fig.~\ref{fig64_f}; the extremum at $\theta=\pi/2$ follows from the symmetry, although only the interval up to $\pi/2$ is displayed.

As shown in Fig.~\ref{fig_snr}(a), the bound on the estimation error $C^{S}$ diverges for $|\boldsymbol{E}| \rightarrow 0$, since the ground state becomes degenerate.
Moreover, the bound $C^{S}$ increases also for sufficiently large values of $|\boldsymbol{E}|$.
Overall, this leads to the existence of an optimal working point, this is also confirmed by the behavior of $\overline{\text{mSNR}}$, shown in Fig.~\ref{fig_snr}(a).
To study the optimal operating point, we fixed $\theta$ and analyzed $E_{\text{O}}$ as a function of $|\boldsymbol{B}|$, where $E_{\text{O}}$ is the value of $| \boldsymbol{E}|$ that maximizes the $\overline{\text{mSNR}}$.
The optimal value $E_{\text{O}}$ is plotted in Fig.~\ref{fig_snr}(b), where it can be observed to be a monotonically increasing with $B$ in the considered range.

Regarding optimal measurements, we notice that the probe $|E_0\rangle$ satisfies the condition Eq.~\eqref{RPDcond}, i.e. both the state vector and its derivatives are described by real coefficient on the basis of energy eigenstates.
Consequently, any PVM satisfying condition Eq.~\eqref{RPMcond}, i.e. defined by basis vectors with real components, also satisfies the optimality condition in Eq.~\eqref{pezzeort}.
This means that an optimal measurement giving a CFIM that attains the QFIM can always be found.

\subsection{Probes at thermal equilibrium}
Next, we consider
the Gibbes thermal state of
the complete Hamiltonian $\hat{H} = \hat{H}(\lambda_1, \lambda_2 , \lambda_3)$ as the probe.
It takes the form
$\rho = \exp[-\beta \hat{H}- \log{Z}]$, with $ Z =\Tr[e^{- \beta \hat{H}}]$ and $\beta = \frac{1}{k_B T}$, where $k_B$ is the Boltzmann constant.
We introduce the operator $\hat{G} = -\beta \hat{H} - \log{Z} \hat{\id} $ with eigenvalues $\{g_n\}$, such that $ \hat{G} | E_n \rangle = g_n |E_n \rangle$.
Since this state is in exponential form, we can exploit the results of Ref.~\cite{Jiang_2014} to compute the QFI.
These are summarized in Appendix~\ref{expo_states}.
In particular, applying Eq.~\eqref{QFIMexpo}, we obtain: 
 \begin{align}
      \notag Q_{\mu \nu} = &   \sum_{m,n} \frac{e^{-\beta E_n}}{Z} \chi_{mn}^2 \Se \bigg[ \langle E_n |\beta \partial_{\mu} \hat{H} + \frac{\partial_{\mu}Z}{Z} |  E_m \rangle \times  \\ &  \times \langle  E_m|\beta \partial_{\nu} \hat{H} + \frac{\partial_{\nu}Z}{Z} | E_n \rangle \bigg]\,,
\end{align}
where $\chi_{m,n}$ are coefficients that depend on the Hamiltonian spectrum, defined in Eq.~\eqref{chi}.
Once again, we obtained an expression that does not require numerical evaluation of the derivatives of the eigenstates.

\begin{figure*}
    \centering
    \includegraphics[width = 17cm]{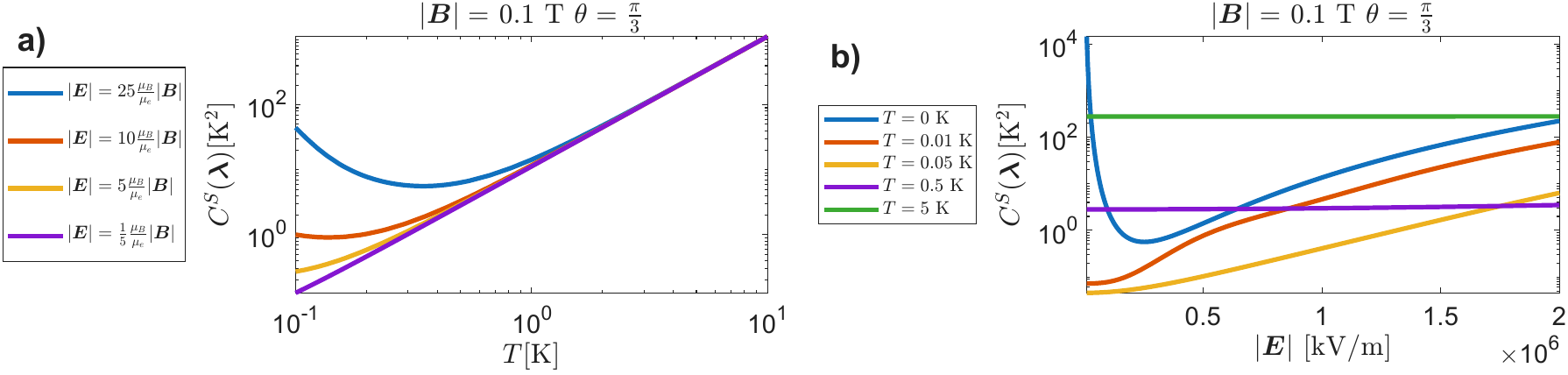}
     \caption{
     (a) Scalar bound on the total estimation error $C^S( \boldsymbol{\lambda}$), in units of kelvin${}^2\,[\text{K}^2]$, for the set of parameters $(\lambda_1,\lambda_2, \lambda_3)$ for a static thermal probe, as a function of the temperature $T$ for fixed  configurations of $| \boldsymbol{B}|$ and $| \boldsymbol{E}|$.
     (b) $C^S( \boldsymbol{\lambda})$ as a function of the magnitude of the electric field $| \boldsymbol{E}|$ and for fixed values of $| \boldsymbol{B}|$, $\theta$ and $T$.
     }
     \label{fig_gibbs}
\end{figure*}

Within the effective Hamiltonian, the formal low-temperature expansion around the ground-state limit gives~\footnote{As discussed in Sec.~\ref{OH}, this low-temperature expansion should be interpreted as an expansion within the effective Hamiltonian.
The finite-temperature numerical results shown in Fig.~\ref{fig_gibbs}, including the curves at $T=10\,\mathrm{mK}$, are above the quoted $5\,\mathrm{mK}$ validity scale.}:
\begin{align} \label{expansion_smallT}
    Q_{\mu \nu}  \notag = & 4 \sum_{n= 1}^7 \frac{\tanh^2{\frac{\beta\,\delta E_n}{2}}}{\delta E_n^2} \, \Se[\langle E_0 |\partial_{\mu} \hat{H}| E_n \rangle\langle E_n |\partial_{\nu} \hat{H}| E_0 \rangle] \\ \notag  =&  Q^{GS}_{\mu \nu}-16\sum_{n= 1}^7\frac{e^{-\beta\,\delta E_n}}{\delta E_n^2}\Se[\langle E_0 |\partial_{\mu} \hat{H}| E_n \rangle \times \\ & \times \langle E_n |\partial_{\nu} \hat{H}| E_0 \rangle]+ O\left( \frac{e^{-2\beta\,\delta E_1}}{\delta E_1^2}\right)\,.
 \end{align}

The expansion in Eq.~\eqref{expansion_smallT} shows that the absolute value of all the matrix elements of the QFIM decreases for increasing temperature in the formal low-temperature regime of the effective model.
This means that all the diagonal elements, i.e. the single-parameter sensitivities for the different parameters, decrease with $T$.
However, diagonal and off-diagonal elements have a different dependence on $T$.
Reducing the magnitude of the off-diagonal elements of the QFIM relatively to the diagonal elements, i.e. reducing correlations between the parameters, generally decreases $C^S$.
Intriguingly, this produces a nontrivial effect on the multiparameter error: it can be seen from Fig.~\ref{fig_gibbs}(b) that $C^S( \boldsymbol{\lambda})$ does not necessarily increase with temperature, and often a minimum for $T>0$ may be observed.
By increasing temperatures, the thermal state eventually becomes too mixed, i.e., it approaches the maximally mixed state, for which any measurement will result in a flat probability distribution.
Thus, there is a tradeoff between a reduced multiparameter error obtained by reducing the correlations, and the state becoming more mixed.
An analogous behaviour has recently been highlighted in the context of critical quantum metrology~\cite{Mihailescu2025c}.


Similar conclusions can be obtained looking at Fig.~\ref{fig_gibbs}(a) explicitly showing that, for small values of $T$, the precision is enhanced for any values of $|\boldsymbol{E}|$ with respect to the ground state.
However, increasing the temperature further, the precision is worsened and, as expected, the value of $C^{S}$ tends to become independent from the value of $|\boldsymbol{E}|$.

Regarding the attainability of the HCRB, the WCC holds for this  class of states.
Indeed, let $\rho = \sum_m e^{g_m} |E_m \rangle\langle E_m|$ and $G_{nm}^{\mu} = \langle E_n |\partial_{\mu} G |E_m \rangle = G_{mn}^{\mu} $, since $G$ is real and hermitian, we have:
\begin{align}
    & \notag -2iD_{\mu \nu} = \Tr\big[\rho [\hat{L}_{\mu}, \hat{L}_{\nu}]\big] \\ &= \sum_{mn} e^{g_m} \chi_{mn}^2  \big[G_{mn}^{\mu} G_{nm}^{\nu}- G_{mn}^{\nu} G_{nm}^{\mu}\big] = 0\,.
\end{align}
However, concerning optimal measurements in the single-copy scenario, the equality between the FIM and the QFIM is not achievable exactly since the SLDs do not satisfy the PCC.
Anyway, the fact that the WCC holds ensures that the QCRB is saturable performing a collective measurement on an asymptotically large number of copies of the probe, as mentioned in Sec.~\ref{multipar_QET}.
Performing such collective and entangling measurements on multiple molecular systems is not yet possible with current technologies, to the best of our knowledge.
Nonetheless, both theoretical and experimental efforts are progressing in entangling molecular systems~\cite{Picard2025,Holland2023,he2023entangled}.

\section{Parameter estimation with dynamical probes}\label{dynamical}

In this section we will focus on dynamical strategies, i.e. using probe states whose probability amplitude over the eigenstates of the Hamiltonian is not fixed.
As in the previous Section we will consider the system to be isolated, where the dynamics is governed by the von Neumann equation: $\dot{\rho_t} = - i [\rho_t, \hat{H}]$.
Assuming a Hamiltonian $\hat{H}$ independent of time and let $\rho_{t = 0} = \rho_0 $, the solution is given by: $\rho_t = e^{-i t \hat{H}} \rho_0  e^{ i t \hat{H}}$.
This scenario is different from the stationary case: information on the parameters is gathered through the temporal evolution of the system, while the initial state of the probe $\rho_0$ does not depend on the parameters.
This opens the possibility to use time as a resource to collect more information about the parameters.
However, the naive idea that, by letting the system evolve freely, the information about the parameters increases monotonically in time is not always true, as we will discuss in the next sections.
This effect is due to the non-commutativity of the Hamiltonian with its derivatives with respect to the parameters.
Nonetheless, a suitable control strategy can restore the total evolution time as a indubitable resource, correcting the effects of non-commutativity and recovering optimal time scaling of the QFI~\cite{Yuan_2015}.

\subsection{Dynamical strategy with aligned fields}

First, we consider again the aligned fields scenario, i.e. the Hamiltonian $\hat{H}_a = \hat{H}(\lambda_1, \lambda_2 = \mu_e | \boldsymbol{E}|, \lambda_3 = 0)$, this time with a dynamical probe.
For a pure initial state $| \psi _0 \rangle$, the QFIM elements are expressed as covariances of the local generators of translations with respect to the parameters $\hat{\mathcal{H}}_{\lambda_\mu} \equiv i \left( \partial_{\lambda_\mu} \hat{U} \right)^\dag \hat{U} $, yielding:
\begin{align}\label{qfiascov}
    & Q_{\mu \nu}  = 4 \, \operatorname{cov}\left( \mathcal{H_{\lambda_\mu}} ,\mathcal{H_{\lambda_\nu}} \right)_{\psi_{0}} \\
    & \equiv  4 \biggl(  
     \frac{1}{2}  \,\langle \psi_0 |\{\mathcal{\hat{H}}_{\lambda_\mu},\mathcal{\hat{H}}_{\lambda_\nu} \} |\psi_0 \rangle  - \langle \psi_0 |\hat{\mathcal{H}}_{\lambda_\mu} | \psi_0 \rangle \langle \psi_0 |\hat{\mathcal{H}}_{\lambda_\nu} |\psi_0 \rangle \biggr). \nonumber
\end{align}
More details are given in Appendix~\ref{app:unitary_metrology}.
For aligned fields, it is possible to express the generators $\hat{\mathcal{H}}_{\lambda_1}$ and $\hat{\mathcal{H}}_{\lambda_2}$ analytically. 
In particular, we have that $\hat{\mathcal{H}}_{\lambda_1} = -\partial_{\lambda_1} \hat{H} t$, and, as expected, for small $t$: $ \hat{\mathcal{H}}_{\lambda_2} = -\partial_{\lambda_2} \hat{H} t + O(t^2)$.
More details and the full explicit expressions are relegated to Appendix~\ref{Hop_analytical}, see Eqs.~\eqref{HopB} and \eqref{HopE}.

We evaluated the QFIM for two different initial states, chosen as possible configurations of the degenerate ground state of the Hamiltonian in absence of fields $\hat{H}(\lambda_1 = 0, \lambda_2 = 0, \lambda_3 =0)$:
\begin{align}
    |\psi_0^A \rangle &= \frac{1}{\sqrt{2}}\left(|e_5\rangle + |e_8 \rangle\right) \\
     |\psi_0^B \rangle &= \frac{1}{2}\left(|e_5\rangle + |e_6 \rangle+|e_7\rangle + |e_8 \rangle\right)\,.
\end{align}
The first state $|\psi_0^A \rangle $ has been chosen because it maximizes the entry of the QFIM relative to $\lambda_1$, while the second is a balanced superposition of the basis in the degenerate minimum-energy subspace in absence of fields.
Explicit expressions of the scalar bound for both states are relegated to Appendix \ref{Hop_analytical}, see Eqs.~\eqref{scalarboundA} and \eqref{scalarboundB}, since they are rather cumbersome.
However, we notice that results are independent of both the phases of the initial state and of $\lambda_1$.

\begin{figure*}
    \centering
    \includegraphics[width = 16cm]{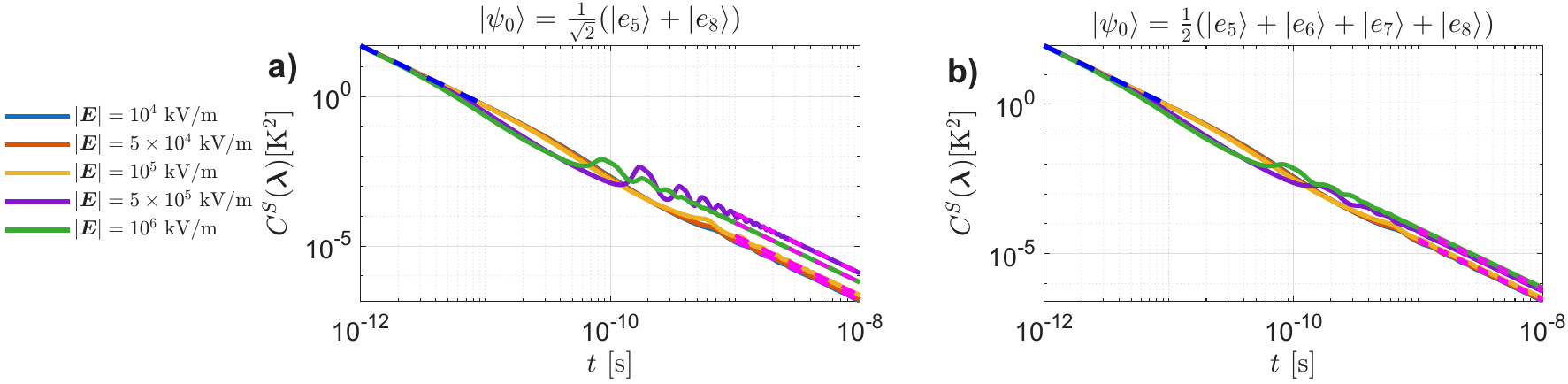}
     \caption{
     Scalar bound on the total estimation error $C^S( \boldsymbol{\lambda})$, in units of kelvin${}^2\,[\text{K}^2]$, for the set of parameters $(\lambda_1,\lambda_2, \lambda_3)$, as a function of the evolution time and for different values of $E$. 
     Two different pure initial states are considered: $|\psi_0^A \rangle = \frac{1}{\sqrt{2}}\left(|e_5\rangle + |e_8 \rangle\right) $ in panel (a), and  $
     |\psi_0^B \rangle = \frac{1}{2}\left(|e_5\rangle + |e_6 \rangle+|e_7\rangle + |e_8 \rangle\right)$ in panel (b).
     We highlight also the small $t$ behavior in Eqs.~\eqref{expsmalltA} and~\eqref{expsmalltB} (dashed blue line) and the large $t$ limits in Eqs.~\eqref{explargetA} and \eqref{explargetB} (dashed magenta lines)}
     \label{aligned}
\end{figure*}
For both states, three different behaviors can be observed from Fig.~\ref{aligned}.
For $t \rightarrow 0$ we observe quadratic time scaling of the QFIM with a prefactor independent of the parameters.
At larger times we have a transition zone in which the scalar bound oscillates, increasing with time in some cases.
Finally, for $t \rightarrow \infty$ we recover again a quadratic time scaling asymptotically, but now with a prefactor that depends on the value of $\lambda_2$.
Explicit expressions for these short and long-time behaviors are reported in Appendix \ref{Hop_analytical}, see Eqs.~\eqref{expsmalltA} and~\eqref{expsmalltB} (dashed blue lines in Fig.~\ref{aligned}) and Eqs.~\eqref{explargetA} and \eqref{explargetB} (dashed magenta lines in Fig.~\ref{aligned}), respectively.
Here ``short'' and ``long'' times are understood with respect to the inverse energy scales entering the effective dynamics, in particular $\Xi^{-1} = \left( \Delta^2+ \frac{4}{25}\lambda_2^2\right )^{-\frac{1}{2}}$ and $\Omega^{-1} = \left(\Delta^2+ \frac{36}{25}\lambda_2^2\right )^{-\frac{1}{2}}$ appearing in the generators of Appendix~\ref{Hop_analytical}.

Finally, it is possible to show by direct computation that $[\mathcal{H}_{\lambda_1},\mathcal{H}_{\lambda_2}]=0$, which implies that the PCC in Eq.~\eqref{eq371} is satisfied.
Consequently, since we are considering pure states, an optimal projective measurement that attains $C^S$ exists.

 \subsection{Dynamical strategy with thermal states}
Next, we consider a dynamical strategy using a thermal state of the free Hamiltonian.
This means evolving an initial state $\rho_0 = \exp\left(-\beta \hat{H}_0- \log\left(Z\right)\right)$, where $\hat{H}_0 = \hat{H}(\boldsymbol{\lambda}=0)$
with the full Hamiltonian $\hat{H}$ in Eq.~\eqref{ham}.
The eigenvalues of $\hat{H}_0$ are $\pm   \frac{ \Delta}{2}$ each four times degenerate, so $Z = \Tr[e^{-\beta \hat{H}_0}] = 8 \cosh\left(\frac{\beta \Delta}{2}\right)$.
Then, since $\hat{H}_0$ is diagonal in the canonical basis its eigenvectors can be chosen to be $\{|e_j\rangle\}$, and, following Eq.~\eqref{QFIMexpl} and Eq.~\eqref{uc},  we obtain :
\begin{align} \label{eq:QiDthermal}
     Q_{\mu \nu}  +i D_{\mu \nu} =& \,  4 \tanh^2\frac{\beta \Delta}{2} \times \\ & \notag \times \bigg[\frac{e^{\frac{\beta \Delta}{2}}}{Z}\sum_{n = 0}^3 \sum_{m = 4}^7   \langle e_n |\hat{\mathcal{H}}_{\mu} | e_m \rangle \langle e_m| \hat{\mathcal{H}}_{\nu}| e_n \rangle  \\&\notag  +\frac{e^{-\frac{\beta \Delta}{2}}}{Z} \sum_{n = 0}^3 \sum_{m = 4}^7 \langle e_m |\hat{\mathcal{H}}_{\mu} | e_n \rangle \langle e_n| \hat{\mathcal{H}}_{\nu}| e_m \rangle \bigg]
     \\ = &\tanh^2\frac{\beta \Delta}{2} \sum_{n = 0}^3 \sum_{m = 4}^7   \langle e_n |\hat{\mathcal{H}}_{\mu} | e_m \rangle \langle e_m| \hat{\mathcal{H}}_{\nu}| e_n \rangle   \,, \notag 
\end{align}
where the matrix elements of  $\hat{\mathcal{H}}_{\mu}$ are reported in Eq.~\eqref{Hop}.
We point out that in Eq.~\eqref{eq:QiDthermal} only the off-diagonal elements of the effective generators are involved.
Noticing that $\partial_{\lambda_1} \hat{H}$ is diagonal in the canonical basis, one can show that the entries of the QFIM relative to $\lambda_1$ vanish in the small $t$ limit and the aligned field scenario, since in these regimes the effective generator is proportional to the partial derivative.

\begin{figure}
    \centering
    \includegraphics[width = 8cm]{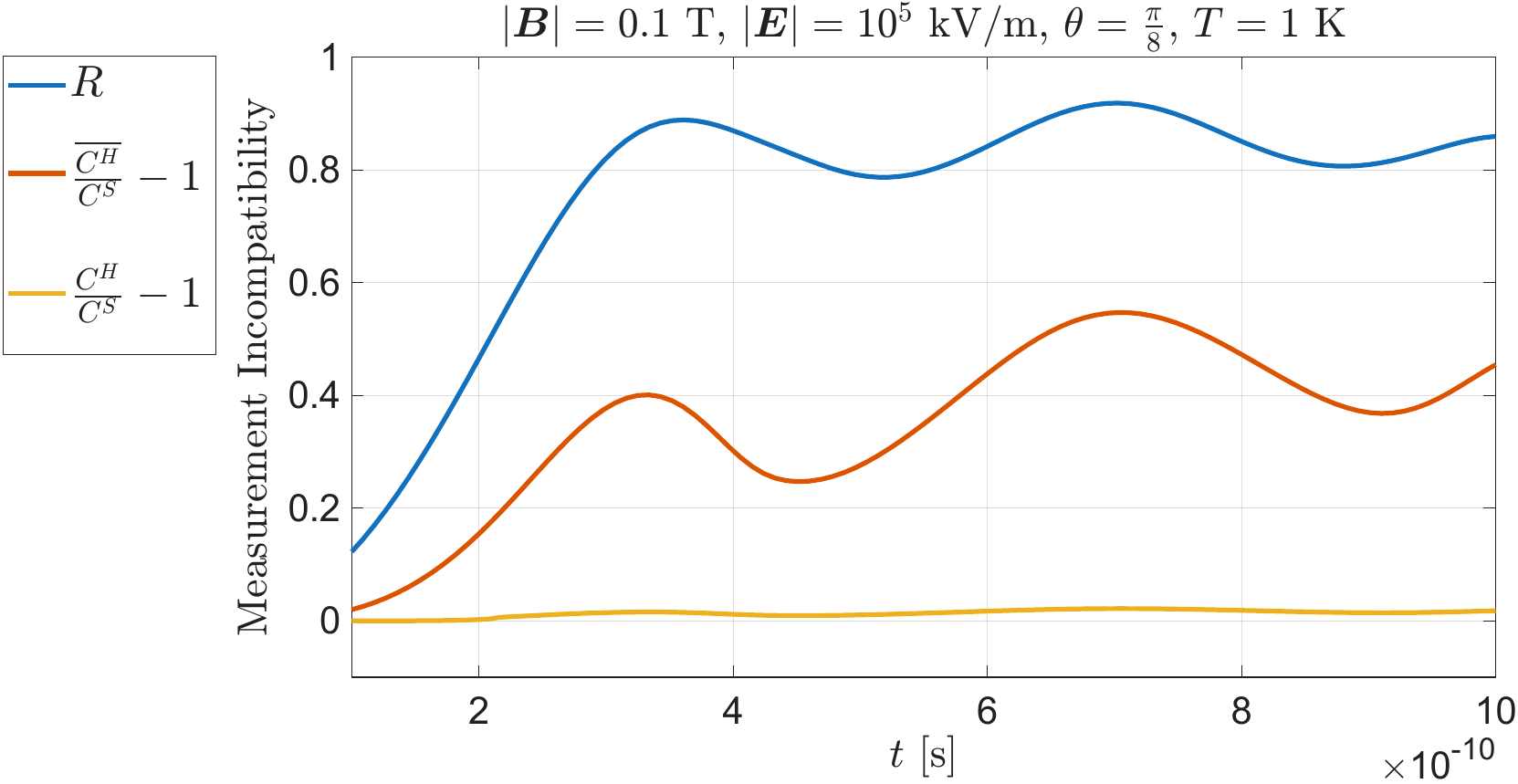}
     \caption{Plot of the asymptotic incompatibility parameter $R$, defined in Eq.~\eqref{R}, the tighter bound $\overline{C^H}/C^S-1$ Eq.\;(\ref{ch_bar}) and $C^H/C^S-1$ which represents the true relative discrepancy between the SLD bound and the HCRB.
     Quantities are plotted as a function of time, for a fixed configuration of fields and temperature.
     It can be easily spotted that neither $R$ nor $\overline{C^H}/C^S-1$ faithfully track the actual incompatibility: even if these quantities are large $C^H$ is almost equal to $C^S$.}
     \label{fighcrb}
\end{figure}

For this model, the mean Uhlmann curvature does not vanish: $D\neq 0$, and thus the incompatibility measure in Eq.~\eqref{R} is greater than zero $R > 0$.
This means that there is a gap between the HCRB and the SLD scalar bound.
However, in Fig.~\ref{fighcrb} we show that both the upper bounds on the HCRB in Eq.~\eqref{R_bound} are far from being tight, and the actual gap between the HCRB and the SLD bound is quite small.
In Fig.~\ref{fighcrb} we have explicitly evaluated the HCRB numerically, 
using the Julia package QuanEstimation~\cite{zhang2022quanestimation,Yu2024} that provides an implementation of the SDP derived in Ref.~\cite{albarelli2019evaluating}.

Thus, we can conclude that the quantity $R$ may not be fully appropriate to assess the incompatibility of a model in a given parametrization of interest~\cite{He_2025,he2025weight}.
It is important to note that $R$ does not depend on the temperature in this scenario.
From its definition in Eq.~\eqref{R}, it is a function of the matrix $Q^{-1} D$, and from Eq.~\eqref{eq:QiDthermal} it is clear the temperature enters only through a multiplicative prefactor that cancels out.


\subsection{Adaptive control schemes}\label{subsec:adaptive}

We have seen that the QFIM for a unitary process with any Hamiltonian dependence on the parameters can be written in Eq.~\eqref{qfiascov} in terms of the effective generators $\hat{\mathcal{H}}_\mu$, as outlined with more details in Appendix~\ref{app:unitary_metrology}.
Crucially, the operators $\hat{\mathcal{H}}_\mu$ are
composed of two parts, as shown explicitly in Eq.~\eqref{Hop}.
The first one is linear in time, and, alone, would lead to quadratic time scaling of the variance.
However, the second part oscillates with time, which could even lead to a time-decreasing QFI in the worst-case scenario.
One would intuitively expect time to be a resource in the sense that information about the parameters should increase with time.
The condition of quadratic time scaling of the information can be restored through the introduction of a sequential control strategy as pointed out for the first time by Yuan and Fung~\cite{Yuan_2015,Yuan_2016}.
The idea is to find a strategy able to linearize the $\hat{\mathcal{H}}_{\mu}$ operators such that the only contribution to the QFI is quadratic in time.

\begin{figure}[ht]
    \centering
    \includegraphics[width = 8cm]{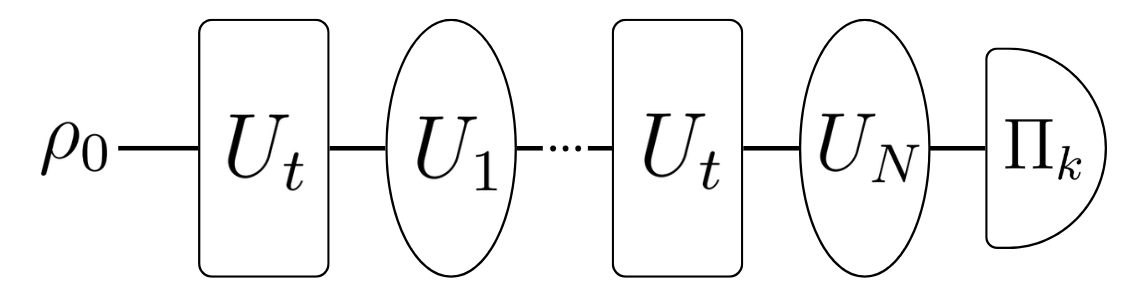}
    \caption{ Pictorial representation of a sequential feedback scheme, where $\hat{U}_t$ is the evolution induced by a Hamiltonian $\hat{H}$ and $\hat{U}_i$, $ i = 1,\dots, N $ are unitary controls}
    \label{fig1}
\end{figure}

The strategy is the following: consider a system that evolves for a time $\tau$ following the unitary dynamics induced by the Hamiltonian $\hat{H}(\boldsymbol{\lambda})$.
Next, we introduce a number $N \gg 1$ of control unitaries $\{\hat{U}_i\}_{i =1\dots N}$ after each interval of time $t = \tau/N \ll \tau$, as depicted in Fig.\;\ref{fig1}.
The total evolution operator becomes: 
\begin{equation}\label{sequential_strategy}
    \hat{U}_{Nt}(\boldsymbol{\lambda}) = \hat{U}_N \hat{U}_t(\boldsymbol{\lambda}) \dots \hat{U}_2 \hat{U}_t(\boldsymbol{\lambda})\hat{U}_1 \hat{U}_t(\boldsymbol{\lambda})\,,
\end{equation}
where \(\hat{U}_t(\boldsymbol{\lambda}) = \exp(-i \hat{H}(\boldsymbol{\lambda}) t)\), \(t = \tau/N\), and \(\hat{U}_i\), \(i = 1, \dots ,N\), are the controls to be implemented that are assumed to employ negligible time.
It is possible to prove that the optimal choice of the control unitaries is: \( \hat{U}_{\text{OC}} = \hat{U}_i = \hat{U}_t^{\dag}(\boldsymbol{\lambda})\) for all \(i < N\) and for arbitrary \(\hat{U}_N\)~\cite{Yuan_2015,Yuan_2016}.
Clearly, $\boldsymbol{\lambda}$ is not known exactly in practice; in this section we assume that its uncertainty is negligible in order to isolate the ideal control mechanism.
The effect of an imperfect estimate of $\boldsymbol{\lambda}$ entering the control operation is addressed in Sec.~\ref{sec:robustness}.
Considering an in initial state $|\psi_0\rangle$, after the evolution given by Eq.~\eqref{sequential_strategy}, in the limit $N \rightarrow \infty$ the QFI reads:
\begin{equation}\label{qfiadaptive}
    Q_{\mu \nu}(\tau)  =4\tau^2\, \operatorname{cov}\left(\partial_{\mu} \hat{H},\partial_{\nu} \hat{H} \right)_{\psi_0 } \,, 
\end{equation}
which, as anticipated, scales quadratically with time.
Formally, this expression amounts to substituting the effective generators appearing in Eq.~\eqref{qfiascov} with their first part, which is linear in time and proportional to the parametric derivatives of the Hamiltonian.
Recently, there have been also experimental implementations of this protocol, both optical~\cite{Hou2019,Hou2020} and with superconducting circuits~\cite{PhysRevLett.132.250204}.

\begin{figure}
    \centering
    \includegraphics[width = 8cm]{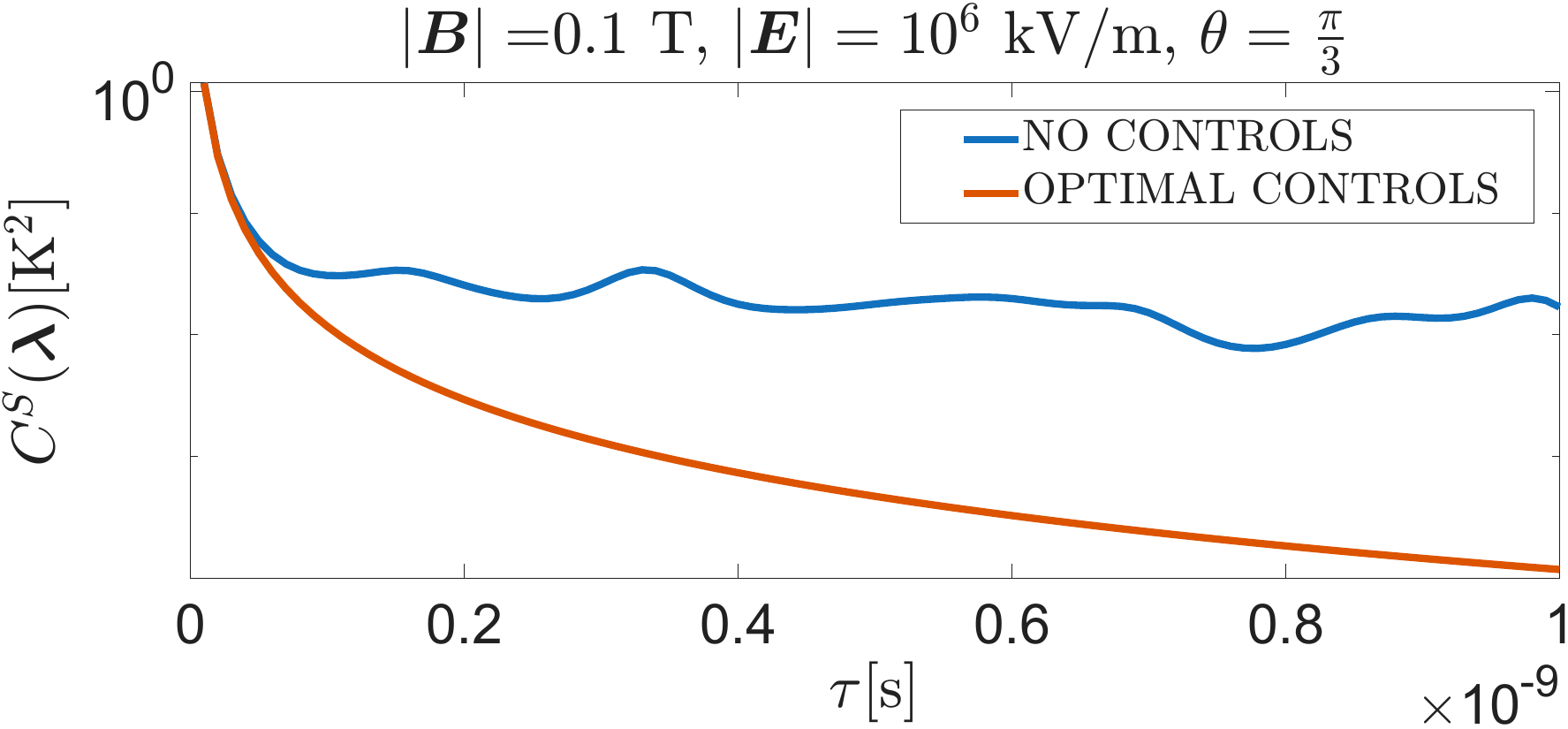}
     \caption{ 
      Scalar bound on the total estimation error $C^S( \boldsymbol{\lambda}$), in units of kelvin${}^2\,[\text{K}^2]$, for fixed values of the parameters and a total evolution time $\tau$.
      This quantity is obtained for the same initial state  $| \psi_0 \rangle $, undergoing an evolution with optimal controls (red line), and without any control (blue line).
      Interestingly, in the uncontrolled scenario (blue line), $C^S(\boldsymbol{\lambda})$ can increase with $\tau$, implying that the information about the parameters decreases with time.}
     \label{figcontrols}
\end{figure}

Going back to our problem of estimating the three parameters $(\lambda_1$, $\lambda_2$, $\lambda_3)$, we seek for the optimal initial state $| \psi_0 \rangle$, minimizing $C^S( \boldsymbol{\lambda}) = \Tr[Q^{-1}]$.
We have numerical evidence that the optimal initial state is:
\begin{equation} \label{opt_probe}
    |\psi_0\rangle =\frac{1}{\sqrt{10}}(2|e_1\rangle + |e_2 \rangle+ |e_3 \rangle+ 2|e_4 \rangle) \,.
\end{equation}
For this probe the QFIM is diagonal in the $N \rightarrow \infty$ limit, with non-zero entries given by: 
\begin{align}
      Q_{\lambda_1\lambda_1} & \label{qfimB}= \frac{592}{125}\tau^2\,, \\
     Q_{\lambda_2\lambda_2} & \label{qfimEz}= \frac{148}{125}\tau^2 \,,\\
      Q_{\lambda_3\lambda_3} & = \frac{4}{125}(19+8 \sqrt{3})\tau^2\,,
      \label{qfimnondiag}
\end{align}
resulting in an optimal value of:
\begin{equation} \label{invqfi}
    \Tr[Q^{-1}] \approx \frac{2.01}{\tau^2}\,.
\end{equation}
Clearly $|\psi_0\rangle$ is independent of the values of the parameters, since the control strategy makes the QFI depend only on the derivatives of the Hamiltonian, which are constant since the Hamiltonian has a linear dpendence on the fields.
While implementing the control operations may be challenging, having a parameter-independent optimal initial state makes the overall strategy more practical.
The advantage obtained by using the optimal control strategy is shown in Fig.~\ref{figcontrols}, where $C^S( \boldsymbol{\lambda})$ is plotted for the free evolution and for the controlled evolution, taking $| \psi_0 \rangle$ Eq.~\eqref{opt_probe} as initial state.
Interestingly, it is possible to observe that exist regions where $C^S( \boldsymbol{\lambda})$ increases with time in the uncontrolled scenario.

We notice that, even if the scalar bound $C^S( \boldsymbol{\lambda})$ does not decrease monotonically in time, it can show a quadratic scaling $\tau^{-2}$ for long times even in absence of control operations, as previously shown for the aligned fields in Fig.~\ref{aligned}.
For suboptimal initial states, it is also possible that the long-time parameter-dependent coefficient of the quadratic scaling is smaller (i.e. better) than the short-time one.
Using a control strategy with optimal initial preparation, the optimal parameter-independent quadratic scaling is maintained throughout the whole evolution.

\subsection{Robustness of the adaptive scheme}
\label{sec:robustness}

Assuming to be able to prepare the optimal probe state $| \psi_0 \rangle$ in Eq.~\eqref{opt_probe} and to implement the control strategy depicted in Fig.~\ref{fig1}, the optimal QFIM will be the one in Eq.~\eqref{qfiadaptive}.
However, in order to implement the optimal control unitaries $\hat{U}_{\text{OC}}(\boldsymbol{\lambda})$ we would need complete knowledge of the true value of the parameters $\boldsymbol{\lambda} = (\lambda_1, \lambda_2, \lambda_3)$.
A more realistic scenario is to consider an estimator $\tilde{\boldsymbol{\lambda}}$ of the parameters $\boldsymbol{\lambda}$ and implement a control unitary $\hat{U}_{\text{OC}}(\tilde{\boldsymbol{\lambda}})$ which is built from the estimator $\tilde{\boldsymbol{\lambda}}$ of the parameters.

Extending the approach of Ref.~\cite{Pang_2014} to multiple parameters \footnote{See also Ref.~\cite{Wei2025b} for a similar approach for multiple parameters, which appeared as a preprint after the initial version of this paper.}, we address the behavior and robustness of the strategy under these nearly optimal controls $\hat{U}_{\text{OC}}(\tilde{\boldsymbol{\lambda}})$, assuming the quantity $\boldsymbol{\eta} = \tilde{\boldsymbol{\lambda}}-\boldsymbol{\lambda}$ to be small.
In order to do that, let us consider a binary POVM~\footnote{This is known as a ``null-measurement'': the outcome corresponding to the projection on $|\psi_0\rangle \langle \psi_0|$ is observed with overwhelming probability, and any deviation from this outcome carries a lot of information on the value of the parameter.
It is optimal in local quantum estimation of pure states, however a rigorous convergence analysis presents some subtleties~\cite{Girotti2024a},  and it is also particularly susceptible to noise in the implementation~\cite{Kurdzialek2023,Albarelli2024b}.}:
\begin{equation} \label{povmtext}
    \left \{ \hat{\Pi}_k \right\} = \{|\psi_0 \rangle \langle \psi_0 |, \hat{\id}-|\psi_0 \rangle \langle \psi_0 | \}\,,
\end{equation}
which is known to be optimal for $\boldsymbol{\eta} = 0$, i.e. the FIM obtained evolving the initial state $|\psi_0\rangle$ in Eq.~\eqref{opt_probe} with the evolution operator given by Eq.~\eqref{sequential_strategy}, implementing the 
optimal controls $\hat{U}_{\text{OC}}(\boldsymbol{\lambda})$ and then performing the measurement in Eq.~\eqref{povmtext} attains the QFIM in Eqs.~\eqref{qfimB}, \eqref{qfimEz} and \eqref{qfimnondiag}.

Expanding the FIM to the second order in $\eta$, we find:
\begin{align}
     F_{\mu \nu} =& \notag 4\tau^2 (K_{\mu \nu}- \Gamma_{\mu} \Gamma_{\nu}) -4\tau^4 \eta^{\rho} \eta^{\sigma}[ K_{\rho \sigma}(K_{\mu \nu} -\Gamma_{\mu } \Gamma_{\nu}) \\ & + K_{\mu \rho } K_{\nu \sigma } - \Gamma_{\rho } \Gamma_{\sigma } \Gamma_{\mu} \Gamma_{\nu}] + O(\eta^4)
\end{align}
where repeated indexes are summed, $\eta^{\mu}$ stands for the $\mu$-th component of the vector $\boldsymbol{\eta}$ and $\Gamma_{\mu} =  \langle \psi_0 |\partial_{\mu}H | \psi_0 \rangle $, $ 
    K_{\mu\nu} = \Se \left[\langle \psi_0 |\partial_{\mu}H \partial_{\nu}H| \psi_0 \rangle \right]$.
More details on these calculations are given in Appendix~\ref{FIexp}. 
Moreover, implementing the probe described in Eq.~\eqref{opt_probe}, we get:
\begin{align} 
  \notag \Tr&[F^{-1}] \\ &\approx \frac{2.01}{\tau^2} +2.63\eta_{\lambda_1}^2  + 0.84  \eta_{\lambda_2}^2 + 0.78 \eta_{\lambda_3}^2 + O(\eta^4)
\end{align}
As expected, at the zeroth order in $\boldsymbol{\eta}$ we recover Eq.~\eqref{invqfi}, meaning that the POVM in Eq.~\eqref{povmtext} would be optimal in the ideal scenario $\boldsymbol{\eta} = 0$, while at the second order a positive correction due to inaccuracy of the estimator.
It is interesting to point out that second order correction does not scale with time, and consequently, can be overcome through large enough evolution times.

\section{Conclusions} \label{conclusions} 
We have addressed the OHM as an effective quantum probe for estimating the direction and intensity of static electric and magnetic fields in configurations that are close to those commonly considered in molecular experiments.
In particular, we have considered stationary strategies, using probes which do not evolve with time, and dynamical ones, where the parameters are encoded onto the probe state via unitary time evolution.
We have evaluated the ultimate bounds to precision according to quantum estimation theory and have sought for optimal realistic strategies approaching those bounds.
Our aim has been to establish quantum-limited benchmarks for this concrete Stark--Zeeman molecular system, rather than to propose a complete experimental sensing architecture.

Concerning stationary probes, we have considered the OHM interacting with aligned fields and have analyzed the metrological performance achievable by measuring the ground state.
Our results indicate that it is only possible to estimate the electric field as a single parameter, while no information can be extracted about the magnetic field.
In this scenario, any measurement described by a real projective POVM is optimal and achieves the scalar QCRB based on the SLDs.
In the general case, where the fields are not parallel, no analytic solution is available and we have solved the problem numerically.
We have investigated the precision achievable using the ground state as a probe, computing the QFIM for different values of the parameters, and
have found that any measurement composed by projectors with real coefficients on the canonical basis is optimal.
We have then considered probes at thermal equilibrium, and have found that $C^H=C^S$, i.e., the ultimate bound to precision may be in principle achieved by collective measurement on an arbitrarily large number of copies of the probe.
However, there is no single-copy measurement strategy such that the FIM is equal to the QFIM, owing to the non-commutativity of the SLDs on the support of the probe.
The thermal-state analysis also illustrates a genuinely multiparameter tradeoff: increasing temperature can populate states that reduce parameter correlations and improve the overall scalar precision in some regimes, despite the simultaneous loss of purity.

We have then investigated dynamical strategies, beginning with the aligned fields scenario.
By analytically evaluating the effective generators and the resulting
QFIM for any evolving state, we have demonstrated that unlike the stationary case, 
the magnetic field can indeed be estimated, and the overall precision increases with the interaction time.
Then, we have considered the general case, with the probe
initially prepared in an equilibrium (thermal) state of the free Hamiltonian.
Contrary to the stationary case, we have found that the PCC does not hold, implying that equality between $C^H$ and $C^S$  cannot be stated a priori.
However, numerical evaluations of $C^H$ have shown that in many configurations $C^S$ and $C^H$ are close to each other, even if the asymptotic incompatibility quantifier $R$ is large.
This confirms that $R$ generally does not track the effects of incompatibility in  multiparameter quantum estimation accurately~\cite{He_2025,he2025weight}.
In particular, asymptotic incompatibility indicators such as $R$ should be interpreted with care, since in this model they do not necessarily quantify the actual finite-copy performance loss in the parametrization of interest.

Finally, we have studied the adaptive scheme of Ref.~\cite{Yuan_2015} to restore the quadratic time scaling of the QFIM, which is lost when the Hamiltonian does not commute with its derivatives with respect to the parameters, as is the case in our system.
This approach uses a feedback strategy to linearize the effective generators.
We optimized the initial state and found, notably, that neither the minimal scalar 
bound on the estimation precision nor the optimal initial state preparation depends on the parameters.
This independence makes the scheme well-suited for practical implementation. Finally, we demonstrated the robustness of the adaptive strategy against imprecise control.

In summary, we have systematically characterized the metrological capabilities of the OHM as a quantum probe for static electric and magnetic fields, comparing stationary and dynamical strategies.
While stationary probes are simpler to implement, they face fundamental limitations—such as the inability to estimate magnetic fields in aligned configurations. Dynamical approaches, by contrast, unlock additional degrees of freedom.
In particular, they generally offer a measurement precision that improves over time, though achieving their optimal realization can be more challenging in practice.
To address this challenge, we introduced and analyzed an adaptive scheme that achieves the optimal quadratic scaling in time and remains robust to control imperfections.
The stationary, freely evolving, and controlled strategies should be viewed as complementary: they use different resources and assumptions, so a direct quantitative ranking would be misleading, but together they identify the mechanisms that enhance or limit joint electric- and magnetic-field estimation.

Overall, these results illustrate the interplay between Hamiltonian structure, measurement optimality, and time evolution in quantum multiparameter estimation, providing insights for the design of practical quantum sensors.

Moreover, this work opens several directions for further research.
On the applied side, we aim to develop a theoretical analysis that moves closer to experimental implementation.
This will require a precise, setup-dependent modelling of the relevant noise sources, going beyond the effective thermal-state approach employed here.
In addition, challenges such as implementing optimal control strategies and collective measurements will need to be addressed in the context of a specific experimental platform.

On the theoretical side, we plan to further investigate the estimation of multiple Hamiltonian parameters from thermal (Gibbs) states.
While general results exist in the single-parameter regime~\cite{Abiuso2025}, a corresponding general theory for the multiparameter case is still lacking.
In this regard, it would be interesting to provide a better understanding of the effect we have highlighted in which increasing the temperature can provide a genuinely multiparameter metrological advantage by reducing the correlations among parameters.
Likewise, control strategies for multiparameter quantum metrology that account for all forms of incompatibility are still under active development~\cite{Hu2024a,Wei2025b}, and become particularly challenging in the presence of noise~\cite{Albarelli_2022}.
Our results for molecular Hamiltonians therefore provide additional motivation to advance the general theory.

\section*{Acknowledgments}
This work has been partially supported by MUR - NextGenerationEU through 
Projects P2022T25TR3-RISQUE, P202222WBL-QWEST, and PE00000023-QMORE. The authors thank Massimo Frigerio for stimulating discussions.

\bibliography{bibliography}

\appendix
\section{Unitary metrology}\label{app:unitary_metrology}
In this section, we present a brief review of the results obtained in recent years about the evaluation of the QFIM in the scenario of unitary metrology~\cite{Liu_2015,Pang_2014}.
Let us assume that the family $\{ \rho_{\boldsymbol{\lambda}} \}$ is generated by the unitary evolution of an initial state $\rho_0 = \sum_n r_n |\psi_n \rangle \langle \psi_n | $ independent of the parameters, i.e. $\rho_{\boldsymbol{\lambda}} = \hat{U} \rho_0 U^{\dag}$, where $\hat{U} = \hat{U}(\boldsymbol{\lambda}) = \exp(-it\hat{H}(\boldsymbol{\lambda}))$ is a unitary operator and $\hat{H}(\boldsymbol{\lambda})$ is Hermitian.
Defining the following set of operators $\hat{\mathcal{H}}_{\mu}$:
\begin{equation} \label{Hopdu}
    \hat{\mathcal{H}}_{\mu} = i\partial_{\mu}\hat{U}^{\dag} \hat{U}\,,
\end{equation}
the QFIM, for a full rank probe in Eq.~\eqref{QFIMexpl}, becomes:
\begin{align}\label{QFI_unitary}
    &  Q_{\mu \nu} = 4 \sum_n r_n\,  \notag \text{cov}\left( \hat{\mathcal{H}}_{\mu} ,\hat{\mathcal{H}}_{\nu} \right)_{\psi_n}\\ & \quad - 8 \sum_{n \neq m} \frac{r_n r_m}{r_n+r_m}\Se\big[\langle \psi_n |\hat{\mathcal{H}}_{\mu} |\psi_m \rangle \langle \psi_m |\hat{\mathcal{H}}_{\nu} |\psi_n \rangle\big]\,,
\end{align}
where the ``covariance'' $\text{cov}( \,\cdot\,, \cdot\,)_{\psi_n}$ on the $n$-th eigenstate is defined as follows.
\begin{align}
    \text{cov}\left( \hat{\mathcal{H}}_{\mu}, \hat{\mathcal{H}}_{\nu} \right)_{\psi_n} = & \notag  \frac{1}{2}\,\langle \psi_n |\{\hat{\mathcal{H}}_{\mu},\hat{\mathcal{H}}_{\nu} \} |\psi_n \rangle \\ & -\langle \psi_n |\hat{\mathcal{H}}_{\mu}| \psi_n \rangle \langle \psi_n |\hat{\mathcal{H}}_{\nu} |\psi_n \rangle\,.
\end{align}
For a pure state model, the QFI in Eq.~\eqref{QFI_unitary} becomes:
\begin{equation}
\label{eq:qfiascov_app}
    Q_{\mu \nu}^{\text{pure}} = 4 \, \text{cov}\left( \hat{\mathcal{H}}_{\mu} ,\hat{\mathcal{H}}_{\nu} \right)_{\psi_{\text{in}}}\,,
\end{equation}
which corresponds to Eq.~\eqref{qfiascov} in the main text.
Then, it is possible to express the operators $\hat{\mathcal{H}}_{\mu}$ in Eq.~\eqref{Hopdu} in a more compact form.
To do that, we define:
\begin{equation}
    \hat{L}^{\text{eff}}_{\mu} = \hat{U}^{\dag} \hat{L}_{\mu} \hat{U} = 2i[\hat{\mathcal{H}}_{\mu}, \rho_0]\,,
\end{equation}
which yields:
\begin{align}\label{QFIMunit}
    Q_{\mu \nu}= &\Tr \bigg[\rho_0  \frac{\hat{L}^{\text{eff}}_{\mu} \hat{L}^{\text{eff}}_{\nu}+ \hat{L}^{\text{eff}}_{\nu} \hat{L}^{\text{eff}}_{\mu}}{2} \bigg]\,.
\end{align}
Then, from the well known equality~\cite{Wilcox:1967zz}:
\begin{equation}\label{der}
    \partial_{\mu} e^{\hat{A}} = \int_0^1 e^{s\hat{A}}(\partial_{\mu} \hat{A}) e^{(1-s)\hat{A}}ds\,,
\end{equation}
we obtain:
\begin{equation}
    \hat{\mathcal{H}}_{\mu} = -\int_0^t e^{is \hat{H}}(\partial_{\mu} \hat{H}) e^{-is \hat{H}}ds \,.
\end{equation}
Finally, defining the superoperator $\Gamma_{\hat{A}} = [\hat{A},\cdot]$ and using the Baker--Campbell--Hausdorff formula~\cite{Wilcox:1967zz}, we obtain:
\begin{equation}\label{Hopexpl}
    \hat{\mathcal{H}}_{\mu} = i \sum_{n= 0}^{\infty} \frac{(it)^{n+1}}{(n+1)!} \Gamma_H^n (\partial_{\mu} \hat{H})\,.
\end{equation}
It is also useful to express $\hat{\mathcal{H}}_{\mu} $ on the eigenbasis of $H$.
Using the fact that:
\begin{equation}
    \hat{U} =\sum _k  e^{-i t E_k}   | E_k \rangle \langle E_k |
\end{equation}
and substituting its derivative in Eq.~\eqref{Hop}, we have:
\begin{align}\label{Hop}
    \hat{\mathcal{H}}_{\mu} =& \notag  -t \bigg[\sum_{k}   \langle E_k |\partial_{\mu} \hat{H}|E_k \rangle | E_k \rangle \langle E_k |\\&  \notag + 4 \sum_{n \neq k} \langle E_n |\partial_{\mu} \hat{H}|E_k \rangle \exp\left(-i t\, \frac{E_n+E_k}{2}\right) \times \\ & \times\sinc\bigg(t\,\frac{E_n-E_k}{2}\bigg)| E_n \rangle \langle E_k | \bigg] \,.
\end{align}
Where we used a generalization of Hellman--Feynman theorem~\cite{PhysRev.56.340}:
\begin{equation} \label{hfth}
      \langle E_n |\partial_{\mu}  \hat{H}|E_k \rangle  = 
      \begin{cases}
          \partial_{\mu}E_n, &\mbox{if } k = n\\
          \langle \partial_{\mu} E_n | E_k \rangle(E_n-E_k)  &\mbox{if } k \neq n \, .
      \end{cases}
\end{equation}

Note that the expression~\eqref{Hop} can be divided in two parts: the first linear in time, which alone would lead to quadratic time scaling of the QFI, as can be easily seen from Eq.~\eqref{eq:qfiascov_app}.
However the second part of Eq.~\eqref{Hop} oscillates with time and could lead, in the worst-case scenario, to a QFI which decreases with time.
This is counterintuitive, since one would expect time to be a resource and so the information about the parameters to be an increasing function of time.
We have addressed this issue in Sec.~\eqref{subsec:adaptive}
using a control scheme to 
linearize $\hat{\mathcal{H}}_{\mu}$ and restore quadratic time scaling of the QFI.

    \section{States in exponential form} \label{expo_states}
In this section, we consider states in the so-called exponential form, i.e., $\rho_0 = e^{\hat{G}_0}$, with $\hat{G}_0$ independent on the parameters, and consequently $\rho= \hat{U} e^{\hat{G}_0} \hat{U}^{\dag} = e^{\hat{U}\hat{G}_0\hat{U}^{\dag}}  = e^{\hat{G}}$.
General result for quantum estimation with this class of states were first presented by Jiang~\cite{Jiang_2014} and we will review them in the following.
Applying Eqs.~\eqref{der} and~\eqref{Hopexpl} to $\partial_{\mu} \rho$ yields:
\begin{equation}\label{eq321}
    \partial_{\mu}\rho \rho^{-1} = \sum_{n = 0}^{\infty} \frac{\Gamma_{\hat{G}}^n(\partial_{\mu}\hat{G})}{(n+1)!}= \bigg[\frac{e^{\Gamma_{\hat{G}}}-1}{\Gamma_{\hat{G}}}\bigg](\partial_{\mu} \hat{G})\,.
\end{equation}
Moreover, from Eq.~\eqref{sldapp} we have:
\begin{align}\label{eq322}
    \partial_{\mu}\rho \rho^{-1} & \notag = \frac{1}{2} (\hat{L}_{\mu}+ e^{\hat{G}} \hat{L}_{\mu} e^{-\hat{G}}) = \frac{1}{2} \bigg[\hat{L}_{\mu}+ \sum_{n = 0}^{\infty} \frac{\Gamma_{\hat{G}}^n( \hat{L}_\mu)}{n!}\bigg] \\ & = \bigg[\frac{e^{\Gamma_{\hat{G}}}+1}{2}\bigg]( \hat{L}_\mu)\,.
\end{align}
Substituting the ansatz:
\begin{equation}
    \hat{L}_\mu = \sum_{n = 0}^{\infty}  f_n \Gamma_{\hat{G}}^n(\partial_{\mu}\hat{G}) = f(\Gamma_{\hat{G}})(\partial_{\mu}\hat{G})
\end{equation}
into Eq.\;(\ref{eq322}), we get:
\begin{align}\label{eq324}
     \partial_{\mu}\rho \rho^{-1} = \frac{e^{\Gamma_{\hat{G}}}+1}{2}f(\Gamma_{\hat{G}})(\partial_{\mu}\hat{G})\,.
\end{align}
Finally, comparing Eqs (\ref{eq321}) and (\ref{eq324}), we obtain:
\begin{equation}
    f(t) = \frac{\tanh\frac{t}{2}}{\frac{t}{2}}\,.
\end{equation}

Now, using the basis of eigenstates of $\hat{G}$, $\hat{G} |g_n \rangle = g_n |g_n \rangle$, the SLDs can be expressed as:
\begin{equation}\label{effectivesld}
    \langle g_n | \hat{L}^{\text{eff}}_{\mu} |g_m \rangle =\chi_{nm} \langle g_n |\partial_{\mu} \hat{G} |g_m \rangle \,,
\end{equation}
and the QFIM is obtained from Eq.~\eqref{QFIMunit}:
\begin{align}\label{QFIMexpo}
     Q_{\mu \nu} = & \notag \sum_{n \ge m} (e^{g_m} +e^{g_m}) \chi_{mn}^2 \times \\ & \times \Se\big[ \langle g_n |\partial_{\mu} \hat{G} |g_m \rangle  \langle g_m |\partial_{\nu} \hat{G} |g_n \rangle \big]\,.
\end{align}
with
\begin{equation} \label{chi}
     \chi_{nm}  = 
      \begin{cases}
          \frac{\tanh \frac{g_n-g_m}{2}}{\frac{g_n-g_m}{2}}, &\mbox{if } m \neq n\\
          1  &\mbox{if } m =  n
      \end{cases}
\end{equation}
If the initial state does not evolve, i.e. $[\rho, H] = 0$, then $\hat{G}= \hat{G}_0$ and the expression~\eqref{QFIMexpo} is definitive.
Otherwise:
\begin{align}
     \langle g_n| \partial_{\mu}\hat{G} |g_m \rangle & \notag =  \langle g_n|\partial_{\mu}\hat{U} \hat{G}_0 \hat{U}^{\dag}|g_m \rangle + \langle g_n| \hat{U} \hat{G}_0\partial_{\mu} \hat{U}^{\dag}|g_m \rangle \\ &= i(g_n- g_m) \langle \psi_n | \hat{\mathcal{H}}_{\mu} | \psi_m \rangle\,,
\end{align}
where $| \psi_m \rangle = \hat{U}^{\dag} | g_m \rangle$ are the eigenstates of $\hat{G}_0$ and $g_n$ are the eigenvalues of both $\hat{G}$ and $\hat{G}_0$, since unitary transformations do not change eigenvalues. Consequently, the QFIM becomes:
\begin{align}
     Q_{\mu \nu} = & \notag 4 \sum_{n >m} (e^{g_m}+e^{g_m})\tanh^2{\frac{g_n-g_m}{2}} \times \\ & \times \Se \big[ \langle \psi_n |\hat{\mathcal{H}}_{\mu} |\psi_m \rangle  \langle \psi_m |\hat{\mathcal{H}}_{\nu} |\psi_n \rangle \big]\,.
\end{align}
    
\section{Effective generators  ${\cal H}$ for the dynamical strategy with aligned fields}
\label{Hop_analytical}

Here, we focus first on the aligned fields scenario, in which $\lambda_3 = 0$ and $\lambda_2 = \mu_e |\boldsymbol{E}|$.
The Hamiltonian Eq.~\eqref{ham} becomes:
\begin{align}\label{hamaligned}
   \hat{H}_a =& \notag  -\Delta\, \hat{\mathcal{T}}_{300}-\frac{8}{5}\lambda_1\, \hat{\mathcal{T}}_{030} -\frac{4}{5}\lambda_1\, \hat{\mathcal{T}}_{003}\\&+\frac{4}{5} \lambda_2\,\hat{\mathcal{T}}_{130}+\frac{2}{5} \lambda_2\, \hat{\mathcal{T}}_{103}\,,
\end{align}
It is trivial to see from Eq.~\eqref{hamaligned} that $[H, \partial_{\lambda_1} H] = 0$.
Then, from Eq.~\eqref{Hopexpl}, we have:
\begin{equation}\label{HopB}
    \mathcal{H}_{\lambda_1} = \frac{8}{5}t\, \hat{\mathcal{T}}_{030} +\frac{4}{5}t\, \hat{\mathcal{T}}_{003}
\end{equation}
Regarding \(\mathcal{H}_E\) we have:
\begin{equation}
    \partial_{\lambda_2}H = \frac{4}{5}\, \hat{\mathcal{T}}_{130}+ \frac{2}{5}\, \hat{\mathcal{T}}_{103}\,.
\end{equation}
At this point, it is straightforward to find the commutation rules.
We have:
\begin{align}\label{mult}
    & \notag \hat{\mathcal{T}}_{ijk}\,\hat{\mathcal{T}}_{lmn} = \frac{1}{4}\left(\sigma_i \otimes \sigma_j \otimes \sigma_k\right)\left(\sigma_l \otimes \sigma_m \otimes \sigma_n\right) \\&= \notag \frac{1}{2}\big[-i \varepsilon_{ila}\varepsilon_{jmb}\varepsilon_{knc} \,\hat{\mathcal{T}}_{abc}-\varepsilon_{ila}\varepsilon_{jmb}\delta_{kn}\,\hat{\mathcal{T}}_{ab0}\\&  \notag  - \varepsilon_{ila }\varepsilon_{knc}\delta_{jm}\,\hat{\mathcal{T}}_{a0c}-\varepsilon_{jmb}\varepsilon_{knc}\delta_{il}\,\hat{\mathcal{T}}_{0bc}+i \varepsilon_{knc}\delta_{il}\delta_{jm}\,\hat{\mathcal{T}}_{00c}\\&   +i \varepsilon_{jmb}\delta_{il}\delta_{kn}\,\hat{\mathcal{T}}_{0b0}+i \varepsilon_{ila}\delta_{jm}\delta_{kn}\,\hat{\mathcal{T}}_{a00}+ \delta_{il}\delta_{jm}\delta_{kn}\,\hat{\mathcal{T}}_{000}  \big]\,,
\end{align}
for \(i,j,k, l,m,n \neq 0\).
Calculating the commutator, the symmetric parts of Eq.~\eqref{mult} vanish:
\begin{align}\label{commrul}
    [\hat{\mathcal{T}}_{ijk},\hat{\mathcal{T}}_{lmn}] & =  \notag  -i \varepsilon_{ila}\varepsilon_{jmb}\varepsilon_{knc} \,\hat{\mathcal{T}}_{abc}+i \varepsilon_{knc}\delta_{il}\delta_{jm}\,\hat{\mathcal{T}}_{00c} \\ & +i \varepsilon_{jmb}\delta_{il}\delta_{kn}\,\hat{\mathcal{T}}_{0b0} +i \varepsilon_{ila}\delta_{jm}\delta_{kn}\,\hat{\mathcal{T}}_{a00}\,.
\end{align}
If one or more indices are zero, it is sufficient to collect the product of the identity for the corresponding Pauli matrix exploiting the associativity of the tensor product and then use the commutation rules on the remaining space.
Indeed:
\begin{equation}
    \hat{\mathcal{T}}_{0nk}\,\hat{\mathcal{T}}_{abc} = \sigma_a \otimes\big[ (\sigma_n  \otimes \sigma_k)(\sigma_b  \otimes \sigma_c)\big]\,.
\end{equation}
As a result, performing the following substitutions: $\Xi = \sqrt{\Delta^2+ \frac{4}{25}\lambda_2^2},\Omega = \sqrt{\Delta^2+ \frac{36}{25}\lambda_2^2}$
and using Eq.~\eqref{Hopexpl} we obtain:
\begin{align}\label{HopE}
    &\hat{\mathcal{H}}_{\lambda_2}
     = \notag-\partial_{\lambda_2} \hat{H} t  \notag  +\frac{4}{5} \Delta\bigg[\left(\hat{\mathcal{T}}_{230}-\hat{\mathcal{T}}_{203}\right)\mathsf{C}_t\left(\Xi\right)\\& \quad + 3\left(\hat{\mathcal{T}}_{230}+\hat{\mathcal{T}}_{203}\right)\mathsf{C}_t\left(\Omega\right) \notag - \Delta\big[\left(\hat{\mathcal{T}}_{130}-\hat{\mathcal{T}}_{103}\right)\mathsf{S}_t\left(\Xi\right) \\& \quad \notag + 3\left(\hat{\mathcal{T}}_{130}+\hat{\mathcal{T}}_{103}\right)\mathsf{S}_t\left(\Omega\right)\big]   +\frac{2}{5} \lambda_2\big[\left(\hat{\mathcal{T}}_{333}-\hat{\mathcal{T}}_{330}\right)\mathsf{S}_t\left(\Xi\right) \\ & \quad - 9\left(\hat{\mathcal{T}}_{333}+\hat{\mathcal{T}}_{330}\right)\mathsf{S}_t\left(\Omega\right) \big]\bigg] 
\end{align}
where $\mathsf{C}_t\left(\Xi\right) = \frac{\cos{\Xi t}-1}{\Xi^2}$ and $ \mathsf{S}_t\left(\Xi\right) = \frac{\sin{\Xi t}-\Xi t}{\Xi ^3}$.
As expected, for small $t$, $ \mathcal{H}_{\lambda_2} = -\partial_{\lambda_2}H t + O(t^2)$.
It is possible now to evaluate the QFIM using Eq.~\eqref{eq:qfiascov_app}.
As an example, we take as initial states two possible configurations of the degenerate ground state of $H(\lambda_1=0, \lambda_2 = 0,\lambda_3 = 0)$:
\begin{align}
    |\psi_0^A \rangle &= \frac{1}{\sqrt{2}}\left(|e_5\rangle + |e_8 \rangle\right) \\
     |\psi_0^B \rangle &= \frac{1}{2}\left(|e_5\rangle + |e_6 \rangle+|e_7\rangle + |e_8 \rangle\right)\,.
\end{align}
The first state, $ |\psi_0^A \rangle$, has been chosen because it maximizes the QFIM entry relative to $\lambda_1$ while the second is a ``flat'' state.
The resulting scalar bounds are:
\begin{widetext}
\begin{equation}\label{scalarboundA}
    \Tr\left[Q^{-1}\left(|\psi_0^A \rangle\right)\right] = \frac{225 \left(8 \Delta ^2 \left(2 \mathsf{C}_t \left(\Omega\right)^2+\mathsf{S}_t \left(\Omega\right) t\right)+16 \Delta ^4 \mathsf{S}_t \left(\Omega\right)^2+5
   t^2\right)+16 \Delta ^2 \lambda_2^2 (\mathsf{S}_t \left(\Xi\right)-9 \mathsf{S}_t \left(\Omega\right))^2}{1296 t^2 \left(16
   \mathsf{C}_t \left(\Omega\right)^2 \Delta ^2+\left(4 \Delta ^2 \mathsf{S}_t \left(\Omega\right)+t\right)^2\right)}
\end{equation}
\begin{equation}\label{scalarboundB}
        \Tr\left[Q^{-1}\left(|\psi_0^B \rangle\right)\right] = \frac{625 \left(\kappa+25 t^2\right)+800 \Delta ^2
   \lambda_2^2 \left(\mathsf{S}_t \left(\Xi\right)^2+81 \mathsf{S}_t \left(\Omega\right)^2\right)}{16 t^2 \left(125 \left(\kappa+5 t^2\right)+96 \Delta ^2 \lambda_2^2 \left(\mathsf{S}_t\left(\Xi \right)^2+12 \mathsf{S}_t \left(\Xi\right) \mathsf{S}_t \left(\Omega\right)+81 \mathsf{S}_t \left(\Omega\right)^2\right)\right)}\,,
\end{equation}
\end{widetext}
where:
\begin{align}
    \kappa = & \notag 8 \Delta ^2 \left(\mathsf{C}_t \left(\Xi\right)^2+9 \mathsf{C}_t \left(\Omega\right)^2+\Delta ^2 \left(\mathsf{S}_t \left(\Xi\right)^2+9
   \mathsf{S}_t \left(\Omega\right)^2\right)\right) \\ & +4 \Delta ^2 t (\mathsf{S}_t \left(\Xi\right)+9 \mathsf{S}_t \left(\Omega\right))
\end{align}
We observed that the QFIM is independent of complex phases of the initial state. Then, it is straightforward to see that the operators $\mathcal{H}_{\lambda_1},\mathcal{H}_{\lambda_2}$, and consequently the QFIM depends only on $\lambda_2$.
It is possible to perform expansions of the two quantities Eqs.~\eqref{scalarboundA} and~\eqref{scalarboundB} in the small $t$ limit: 
\begin{align}\label{expsmalltA}
    \Tr\left[Q^{-1}\left(|\psi_0^A \rangle\right)\right] &= \frac{125}{144}\frac{1}{t^2}+ O(1)\\
    \Tr\left[Q^{-1}\left(|\psi_0^B \rangle\right)\right] &= \frac{25}{16}\frac{1}{t^2}+ O(1)\,, \label{expsmalltB}
\end{align}
resulting, as expected, in quadratic time scaling of the QFIM with the prefactor independent of the parameters.
Regarding the large $t$ limit, we have:
\begin{widetext}
\begin{align}\label{explargetA}
    \Tr\left[Q^{-1}\left(|\psi_0^A \rangle\right)\right]  &= \frac{125 \left(9140625 \Delta ^8+8150000 \Delta ^6
   \lambda_2^2+8172000 \Delta ^4
   \lambda_2^4+2384640 \Delta ^2
   \lambda_2^6+186624\lambda_2^8\right)}{11664 t^2 \left(-625 \Delta ^4+200
   \Delta ^2 \lambda_2^2+48 \lambda_2^4\right)^2}+ O \left( \frac{1}{t^3} \right)\,,\\
    \Tr\left[Q^{-1}\left(|\psi_0^B \rangle\right)\right] &= \frac{25 \left(25 \Delta ^2+4 \lambda_2^2\right) \left(25 \Delta ^2+36
   \lambda_2^2\right) \left(1625 \Delta
   ^4+1288 \Delta ^2 \lambda_2^2+144
   \lambda_2^4\right)}{16 t^2
   \left(3515625 \Delta ^8+4590000 \Delta ^6
   \lambda_2^2+3340000 \Delta ^4
   \lambda_2^4+495360 \Delta ^2
   \lambda_2^6+20736 \lambda_2^8\right)}+ O \left( \frac{1}{t^3} \right)\,,\label{explargetB}
\end{align}
\end{widetext}
which again comes from a quadratic time scaling of the QFIM, since,in this limit,  $\mathsf{C}_t\left(\Xi\right)\approx 0$ and $ \mathsf{S}_t\left(\Xi\right) \approx -\frac{ t}{\Xi ^2}$, and substituting these values in Eq.~\eqref{HopE}, we obtain that $\mathcal{H}_{\lambda_2}$ scales linearly with time.
However, the difference from the behavior in the small $t$ limit is that, now, the prefactor depends strongly on $\lambda_2$.

\section{FI expansion for the sequential strategy}\label{FIexp}

In this section, we provide further details of the results presented in Sec.~\ref{sec:robustness}, regarding a more realistic situation in which we do not assume to have perfect knowledge of the value of the parameters when implementing the control strategy depicted in Fig.~\ref{fig1}.

We recall the following notation introduced in Sec.~\ref{sec:robustness}:
\begin{align}
    \Gamma_{\mu} &=  \langle \psi_0 |\partial_{\mu} \hat{H} | \psi_0 \rangle \label{gamma}\\ 
    K_{\mu\nu} &= \Se \left[\langle \psi_0 |\partial_{\mu} \hat{H} \partial_{\nu} \hat{H}| \psi_0 \rangle \right]\label{K}\,,
\end{align}
where $|\psi_0\rangle$ is the initial state.
Then, after implementing the evolution Eq.~\eqref{sequential_strategy} with nearly optimal controls, the state becomes:
\begin{align} \label{nearly_opt_evo_state}
    |\xi \rangle =& \notag A \left[ \hat{\id} + i\tau (\hat{H}(\tilde{\boldsymbol{\lambda}})- \hat{H}(\boldsymbol{\lambda})) \right] |\psi_0\rangle +O(\tau^2/N) \\
    = & A[ \hat{\id}+ i\tau \eta^{\rho} \partial_{\rho} \hat{H}] |\psi_0\rangle+O(\tau^2/N)+O(\eta^2)
\end{align}
where repeated indexes are summed and $A$ is the normalization factor:
\begin{equation}
    A = [1+\tau \eta^{\rho} \eta^{\sigma}K_{\rho \sigma}  ]^{-1/2}\,.
\end{equation}
As will be clear later, in order to expand the FIM to second order, is enough to expand the evolved state Eq.~\eqref{nearly_opt_evo_state} to first order.
Performing the measurement Eq.~\eqref{povmtext}, we obtain the following FIM:
\begin{equation}\label{fim_nearly_opt_strat}
    F_{\mu \nu} = 4\frac{ \Se[ \langle \psi_0 |\partial_{\mu} \xi\rangle \langle \xi | \psi_0\rangle ] \Se[ \langle \psi_0 |\partial_{\nu} \xi\rangle \langle \xi | \psi_0\rangle ]}{| \langle \xi | \psi_0 \rangle |^2(1-| \langle \xi | \psi_0 \rangle |^2)}
\end{equation}
consequently, we have :
\begin{equation}
    | \partial_{\mu} \xi \rangle = [ \partial_{\mu} A [ \hat{\id} + i\tau\eta^{\rho} \partial_{\rho} \hat{H} ]- i A \tau \partial_{\mu} \hat{H} ] | \psi_0 \rangle+ O(\eta^2)
\end{equation}
and
\begin{equation}
    \partial_{\mu} A =  A^3 \tau^2 \eta^{\rho} K_{\rho \mu} = \tau^2 \eta^{\rho} K_{\rho \mu}+O(\eta^3) \,.
\end{equation}
The factors of the FIM reads:
\begin{align}\label{gatto}
    \Se[ \langle \psi_0 |\partial_{\mu} \xi\rangle& \notag  \langle \xi | \psi_0\rangle ] 
  \\ &= A\partial_{\mu} A -A^2 \tau^2 \eta^{\rho}\Gamma_{\rho} \Gamma_{\mu}+O(\eta^2)\,,
\end{align}
and
\begin{align}
    | \langle \xi | \psi_0 \rangle |^2 = A^2 (1+ \tau^2 \eta^{\rho} \eta^{\sigma} \Gamma_{\rho} \Gamma_{\sigma})+O(\eta^3)\,.
\end{align}
Note that Eq.~\eqref{gatto} is $O(\eta)$, this implies that the numerator of the FIM in Eq.~\eqref{fim_nearly_opt_strat} is $O(t^2)$ while the denominator is $1+O(t)$, this justifies the expansion to first order in Eq.~\eqref{nearly_opt_evo_state}.
Finally:
\begin{align}
      F_{\mu \nu} = & \notag 4\tau^2 (K_{\mu \nu}- \Gamma_{\mu} \Gamma_{\nu}) -4\tau^4 \eta^{\rho} \eta^{\sigma}[ K_{\rho \sigma}(K_{\mu \nu} -\Gamma_{\mu } \Gamma_{\nu})  \\ &+ K_{\mu \rho } K_{\nu \sigma } - \Gamma_{\rho } \Gamma_{\sigma } \Gamma_{\mu} \Gamma_{\nu}]+O(\eta^4)
\end{align}
As expected, at the zeroth order in $\eta$ we get: $4\tau^2 (K_{\mu \nu}- \Gamma_{\mu} \Gamma_{\nu}) = 4\tau^2 \operatorname{cov}( \partial_{\mu} \hat{H} ,\partial_{\nu} \hat{H} )_{\psi_0}$.

\end{document}